\documentclass[final,5p,twocolumn,times]{elsarticle}
\usepackage{amsmath,amsfonts,amssymb}
\usepackage{graphicx}
\usepackage{subcaption}
\usepackage{comment}
\usepackage{lineno}

\journal{Nuclear Instruments and Methods in Physics Research A}

\begin{document} 

\begin{frontmatter}

\title{The Camera and Readout for the Trinity Demonstrator and the EUSO-SPB2 Cherenkov Telescope}

\author[a]{Mahdi Bagheri}
\author[a]{Srikar Gadamsetty}
\author[a]{Eliza Gazda\corref{cor1}}
\ead{elizagazda@gatech.edu}
\author[c]{Eleanor Judd}
\author[b]{Evgeny Kuznetsov}
\author[a]{A. Nepomuk Otte}
\author[a]{Mathew Potts}
\author[a]{Oscar Romero Matamala\corref{cor1}}
\ead{oromero@gatech.edu}
\author[a]{Noah Shapera}
\author[a]{Joshua Sorell}
\author[a]{Svanik Tandon}
\author[a]{Andrew Wang}

\affiliation[a]{
organization={Georgia Institute of Technology, School of Physics, Center for Relativistic Astrophysics},
                addressline={837 State Street NW}, 
                city={Atlanta},
                state={GA},
                postcode={30332-0430}, 
                country={U.S.A.}}

\affiliation[b]{
organization={University of Alabama in Huntsville, Center for Space Plasma and Aeronomic Research},
addressline={NSSTC, CSPAR, 320 Sparkman Drive},
city={Huntsville},
state={AL},
postcode={35805},
country={U.S.A.}}

\affiliation[c]{
organization={University of California at Berkeley, Space Sciences Laboratory},
addressline={7 Gauss Way},
city={Berkeley},
state={CA},
postcode={94720},
country={U.S.A}}

\cortext[cor1]{Corresponding Authors}



\begin{abstract}
  We developed a modular silicon photomultiplier camera to detect Earth-skimming PeV to EeV tau neutrinos with the imaging atmospheric Cherenkov technique. We built two cameras, a 256-pixel camera with S14161-6050HS SiPMs for the \emph{Trinity} Demonstrator located on Frisco Peak, Utah, and a 512-pixel camera with S14521-6050AN SiPMs for the EUSO-SPB2 Cherenkov Telescope. The front-end electronics are based on the eMUSIC ASIC, and the camera signals are sampled and digitized with the 100\,MS/s and 12-bit AGET system. Both cameras are liquid-cooled. We detail the camera concept and the results from characterizing the SiPMs, bench testing, and calibrating the two cameras. 
  
\end{abstract}




\end{frontmatter}

\section{Introduction}
\label{sect:intro}  
The neutrino sky at very-high (VHE, $>$PeV) and ultra-high (UHE, $>$EeV) energies is still dark. However, IceCube's transformational detection of diffuse astrophysical neutrinos \cite{Aartsen2013}, evidence of two neutrino point sources TXS\,0506+056 \cite{Aartsen2018b,Aartsen2017} and NGC\,1068 \cite{Abbasi2022}, and the galactic plane \cite{Abbasi2023a} at high energies ($>$1TeV) highlight the tremendous potential that neutrinos offer to gain fundamental new insight into the non-thermal universe. Extending neutrino observations to higher energies and opening the VHE/UHE neutrino band will provide us with a unique view of the most extreme cosmic particle accelerators, help us understand cosmic-ray propagation and the evolution of the universe, and allow us to study fundamental neutrino physics and probe new physics beyond the standard model of particle physics at the highest possible energies \cite{Ackermann2022}.

Detecting neutrinos is challenging because interaction cross-sections are extremely small. The much lower neutrino fluxes in the VHE/UHE band compared to the HE band exasperate the problem. Overcoming these challenges requires instruments with orders of magnitude larger detector volumes than IceCube's. Of the different proposed detection techniques \cite{Ackermann2022}, we pursue detecting Earth-skimming tau neutrinos with the imaging atmospheric Cherenkov technique \cite{Fargion2001}.

The Earth-skimming technique is sensitive to $>$PeV tau neutrinos that enter the Earth under a small ($<10^\circ$) angle, undergo charged current interaction, and produce a tau lepton. The tau continues on the trajectory of the neutrino, emerges from the Earth, and decays, starting a massive shower of mostly Cherenkov-light emitting electrons and positrons. A Cherenkov telescope captures some of the light and generates an image of the air shower onto a pixelated focal plane, the camera. 
 
This paper discusses the cameras we have developed for the VHE/UHE neutrino instruments the \emph{Trinity} Demonstrator and the EUSO-SPB2 Cherenkov telescope. \emph{Trinity} is a proposed system of 18 Cherenkov Telescopes \cite{Otte2023a,Otte2019d} on mountaintops. Its first development stage is the \emph{Trinity} Demonstrator, which we deployed on Frisco Peak, UT, in the Summer of 2023 instrumented with the camera described here. The \emph{Trinity} Demonstrator is a one-square meter class Cherenkov telescope with Davis Cotton optics and a $5^\circ \times 5^\circ$ field of view (FoV).
The EUSO-SPB2 long-duration balloon mission is a precursor to the proposed POEMMA mission that aims to detect neutrinos from space by looking at the Earth's limb \cite{Olinto2020}. 
The EUSO-SPB2 balloon flew in the Spring of 2023, and we observed the Earth limb with the Cherenkov telescope for two nights from a float altitude of $\sim35$\,km before it crashed into the Pacific due to a leaking balloon \cite{Gazda2023, Eser2023lck}. The Cherenkov telescope on EUSO-SPB2 was a modified Schmidt optics with a 0.8\,m$^2$ light-collection surface and a $6.4^\circ \times 12.8^\circ$ FoV instrumented with the camera described here. 

We start the paper with a discussion of camera design considerations in Section \ref{sec:design} followed by a description of the modular architecture of the camera in Section \ref{sec:architecture}. A description of the main components of the camera and the readout is given in Section \ref{sec:components}. The cooling system and thermal vacuum testing are described in Section \ref{sec:cooling}. The characterization of the photon detectors are detailed in Section \ref{sec:sensors} and the signal chain in Section \ref{sec:signalchain}. The flatfielding of the camera response is discussed in Section \ref{sec:flatfielding} and the current monitor in Section \ref{sec:currentmonitor}.

\begin{figure*}[!t]
\centering
\includegraphics[angle=0,width=0.7\textwidth]{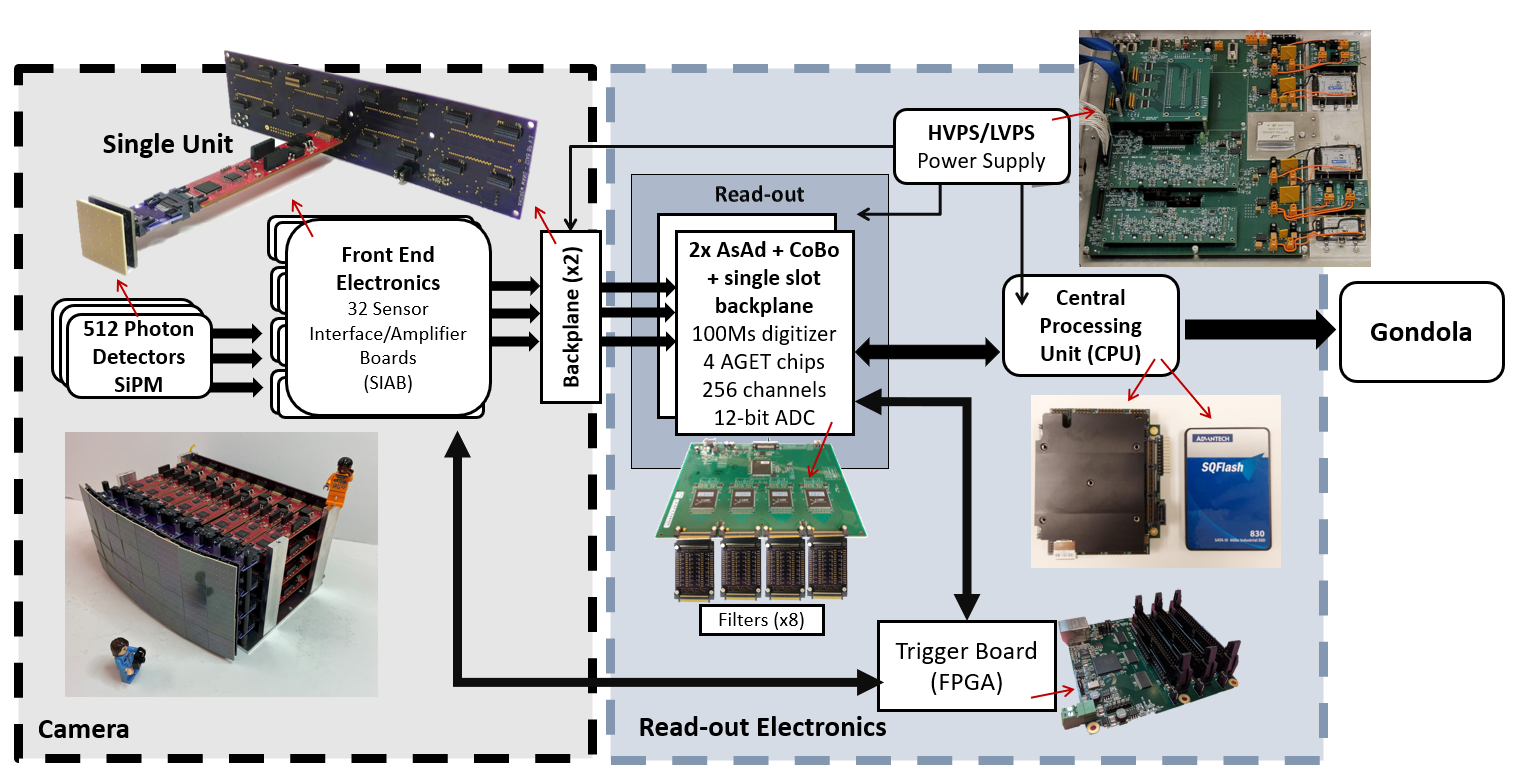}
\caption{Architecture of the camera and readout in the configuration for the EUSO-SPB2 Cherenkov telescope. \label{fig:CameraModules}}
\end{figure*}

\section{Design Considerations\label{sec:design}}

The purpose of a Cherenkov-telescope camera is to detect the faint and fast transient flashes of Cherenkov light emitted by the electrons and positrons in an air shower. A typical Cherenkov flash lasts between a few nanoseconds and a few hundred nanoseconds, depending on the viewing angle of the air shower relative to the air-shower axis. A Cherenkov telescope camera needs to typically detect at least 100 photons (photoelectrons) to guarantee a reliable event reconstruction \cite{Albert2008, Otte2019d}. Furthermore, the few Cherenkov photons must be distinguished from the fluctuations of the ambient light, which is called the night-sky background (NSB) \cite{Benn1999}. These conditions translate into requirements for the analog bandwidth, dynamic range, digitizer sampling speed, maximum acceptable electronic noise, and the trigger of the Cherenkov-telescope camera discussed in the remainder of this section.

For the best separability of the Cherenkov flash from NSB fluctuations, the signal in the readout should have a width of about 10\,ns. However, power constraints and other practical considerations had us design a system with a considerably lower bandwidth, resulting in signals that are 30\,ns full-width at half maximum.  Because fluctuations in the NSB are irreducible, they define the noise floor, and the signal chain must be designed such that any additional noise contributions are much below the NSB fluctuations. On the opposite end of the signal chain's dynamic range, we require a linear response of up to a few hundred photoelectrons, which covers the expected range for most events. For rare extreme events with signals beyond the linear range, the dynamic range can be extended by considering the non-linear response of the signal chain in the analysis.

The angular size of the camera pixels, is driven by the science goals for the \emph{Trinity} Demonstrator and the EUSO-SPB2 Cherenkov telescope. Among other goals, both missions aim to measure backgrounds that mimic air showers from Earth-skimming tau neutrinos. In \emph{Trinity}'s case, we want to record these events with the same $0.3^\circ$ pixel size we plan for the final \emph{Trinity} telescopes. In EUSO-SPB2's case, it was more important to cover a large field-of-view to better study the spatial and temporal characteristics of the NSB, even if it meant that the pixel size would be larger than the $0.083^\circ$ anticipated for POEMMA \cite{Olinto2020}. 

Eventually, we fixed the pixel size for EUSO-SPB2 at $0.4^\circ$ based on the $6^\circ\times12^\circ$ field-of-view of the EUSO-SPB2 Cherenkov telescope optics, the available power, and the projected power consumption per camera channel. Another benefit of this choice is that the required physical pixel size is the same 6\,mm required for \emph{Trinity}. The only difference between the two cameras is the number of pixels, which is 256 pixels for \emph{Trinity} and 512 pixels for EUSO-SPB2 to cover the fields of view of the respective telescopes they instrument. We, therefore, designed the modular camera architecture described in the next section that meets the requirements for both instruments.

The signals of a Cherenkov telescope camera are not continuously recorded. Instead, the trigger electronics continuously scans the camera signals for a potentially interesting signal topology in the camera. If the electronics senses the required topology, the trigger sends a readout command to the digitizer. We provide more details about the trigger topologies required in the \emph{Trinity} Demonstrator and the EUSO-SPB2 Cherenkov telescope in Section \ref{sec:trigger}.

From an operational point of view, the EUSO-SPB2 camera operated in a much more challenging environment. As convective cooling becomes ineffective at 33\,km altitude, the design of the camera cooling was driven by the EUSO-SPB2 requirements and adapted for the \emph{Trinity} camera (see Section \ref{sec:cooling}.)

Based on these design considerations, we devised a modular camera that meets the \emph{Trinity} Demonstrator and EUSO-SPB2 requirements and is described in the following sections.

\section{Top-Level Architecture of the Camera and Readout\label{sec:architecture}}

The top-level architecture is divided into the camera unit and the readout unit. Figure \ref{fig:CameraModules} shows the block diagram of the system and how it breaks down into the two physically separate units. The camera unit is mounted in the telescope's focal plane, and the readout unit is placed in a convenient location. In \emph{Trinity}, the readout resides inside a cabinet on wheels next to the telescope, and in EUSO-SPB2, the readout is inside a box mounted below the telescope.
\begin{figure}[!htb]
    \centering
    \begin{subfigure}[b]{0.9\columnwidth}\includegraphics[angle=0, width=1\columnwidth]{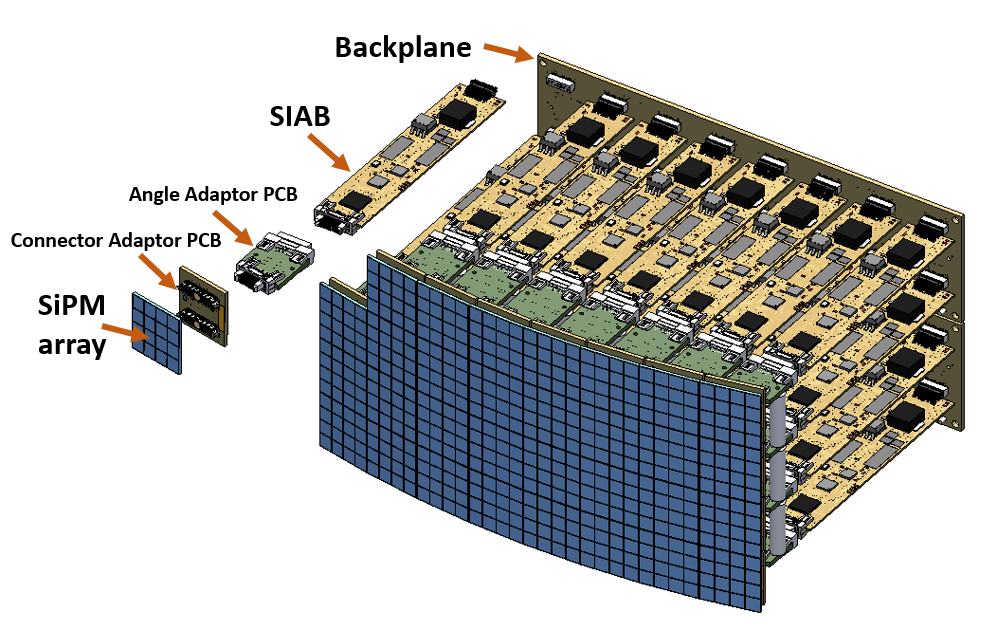}
        \caption{EUSO-SPB2 camera without the cooling system.}
        \label{fig:SPB2CAD}
    \end{subfigure}
    \begin{subfigure}[b]{0.9\columnwidth}
    \centering\includegraphics[angle=0, width=0.9\columnwidth]{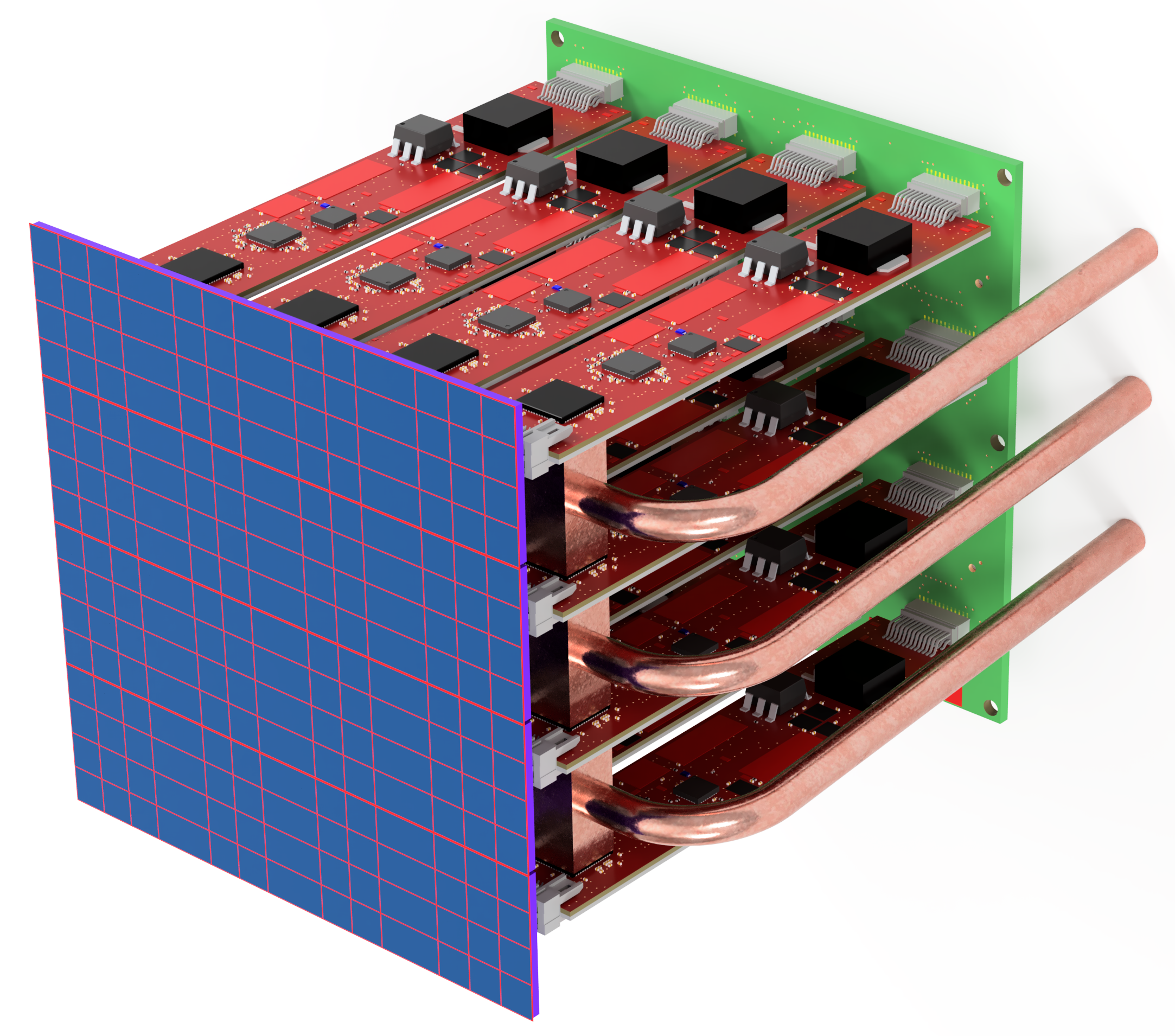}
        \caption{\emph{Trinity} camera including the copper heat pipes for the cooling system.}
        \label{fig:TrinityCAD}
    \end{subfigure}
    \caption{CAD drawings of the EUSO-SPB2 camera and \emph{Trinity} camera. See text for a description.  \label{fig:CameraCAD}}
\end{figure}

\begin{figure}[!htb]
    \centering
    \includegraphics[angle=0, width=0.9\columnwidth]{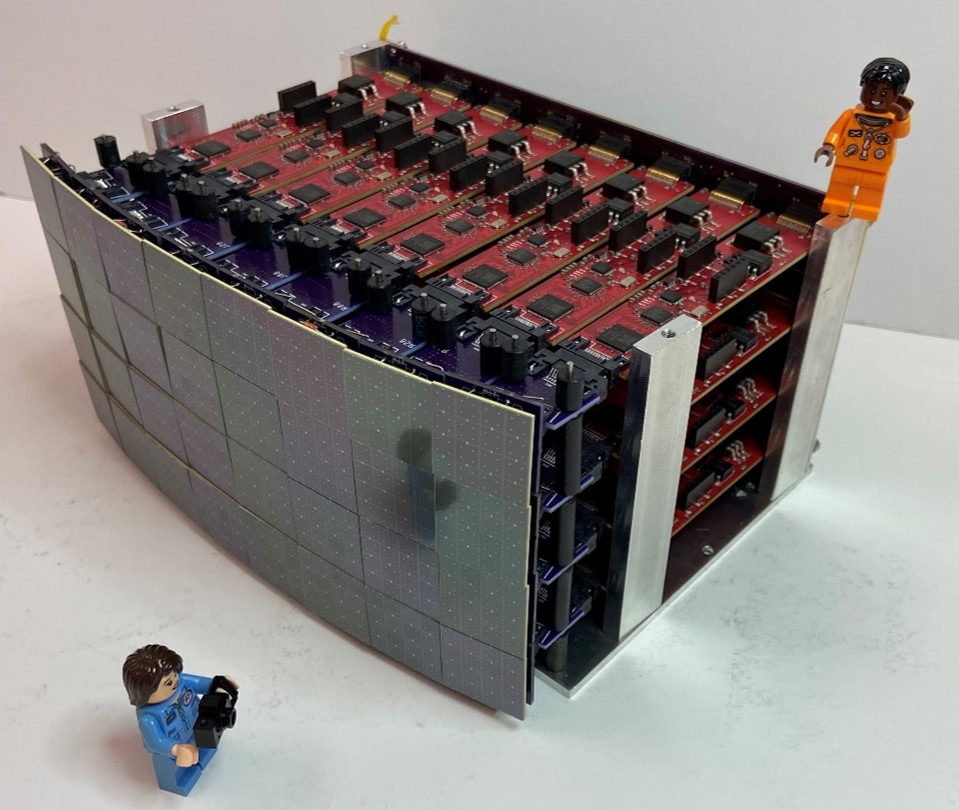}
        \caption{The Euso-SPB2 camera without cooling system and missing top cover. \label{fig:CamPicSPB2}}
\end{figure}

\begin{figure}[!htb]
    \centering
    \includegraphics[angle=0, width=0.9\columnwidth]{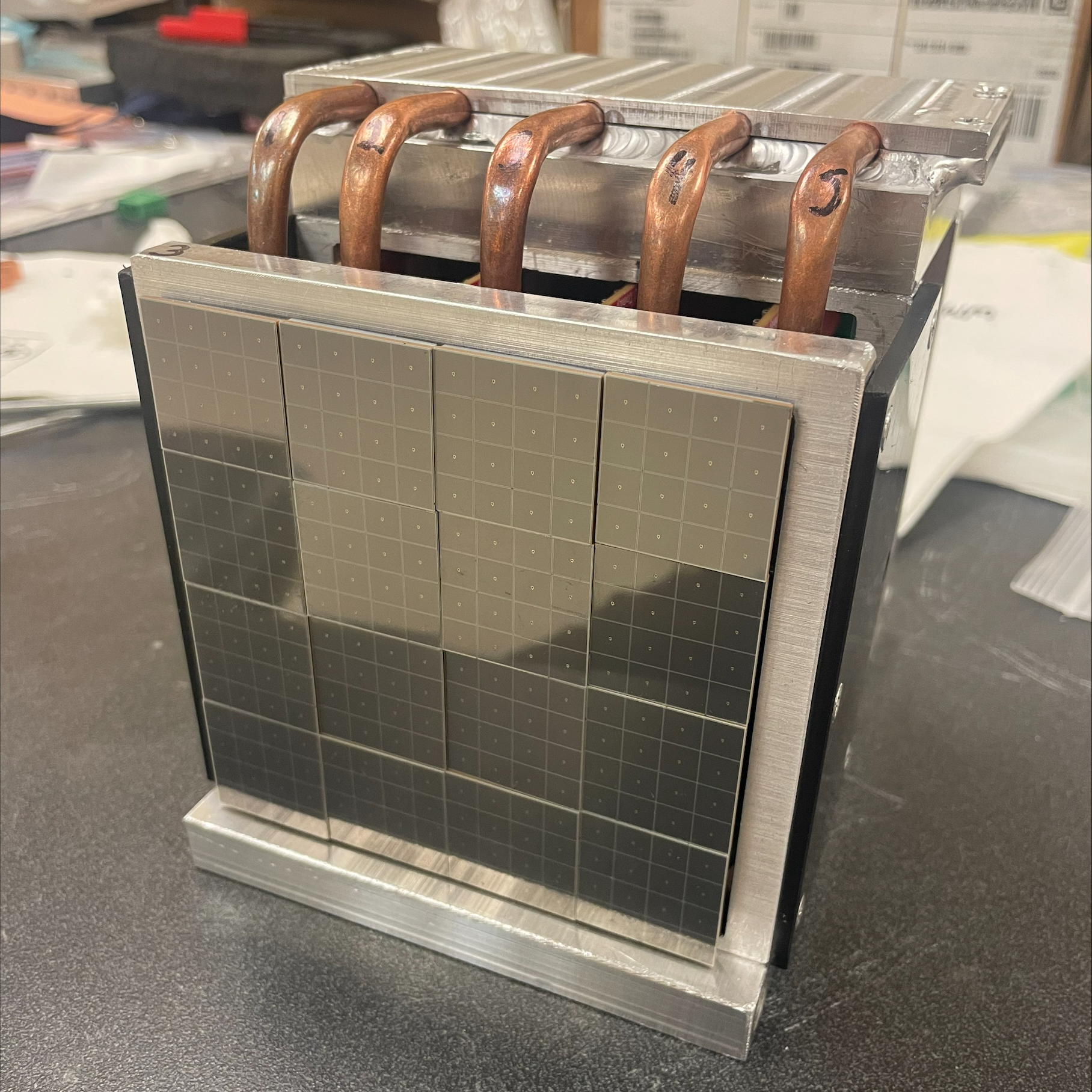}
    \caption{Assembled \emph{Trinity} Demonstrator camera with cooling system.
    \label{fig:CamPicTrinity}}
\end{figure}

\paragraph{The camera unit} 
Figure \ref{fig:CameraCAD} shows CAD drawings of the \emph{Trinity} and EUSO-SPB2 cameras, and Figures \ref{fig:CamPicSPB2} and \ref{fig:CamPicTrinity} show pictures of the two built cameras. The cameras are composed of modules. Figure \ref{fig:CameraCAD} shows the exploded view of one of the modules, which consists of a square matrix of sixteen 6\,mm $\times$ 6\,mm silicon photomultipliers (SiPM). Each SiPM constitutes one camera pixel. The SiPM matrix attaches to the front-end electronics board, the Sensor Interface and Amplifier Board (SIABs). To approximate the curved focal plane of the EUSO-SPB2 optics, an angled interface board is inserted between the SiPM matrix and the SIAB. The focal plane in the \emph{Trinity} Demonstrator is flat, and the interface board is unnecessary and not used.

The 16 modules of the \emph{Trinity} camera and the 32 modules of the EUSO-SPB2 camera  (see Figure \ref{fig:CameraCAD}) are inserted into the camera backplane, which is custom-designed for each camera to accommodate the different number of camera modules. The backplanes provide a mechanical mount point for the modules and the electrical interface to the readout, the trigger, and power.

\begin{figure}[htb]
\includegraphics*[angle=0,width=.9\columnwidth]{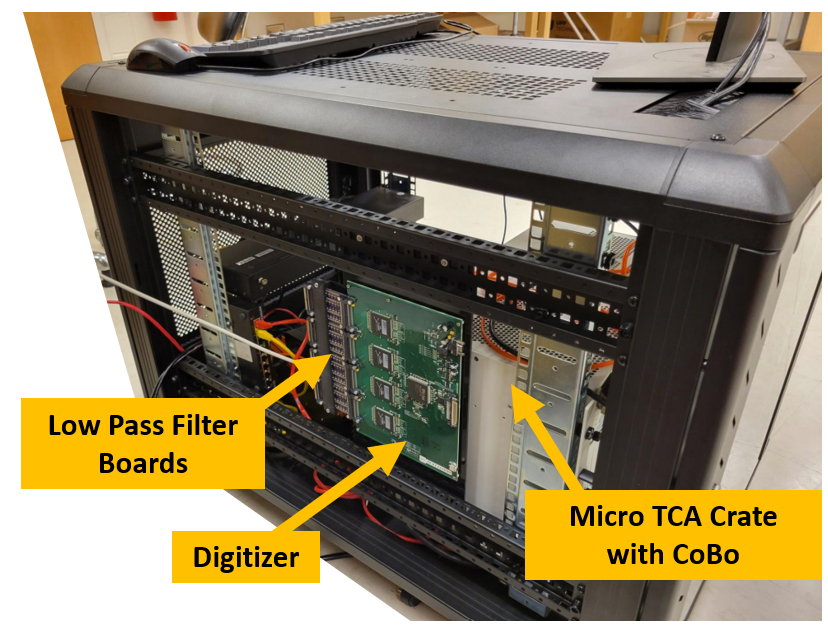}
\caption{The readout unit for the \emph{Trinity} Demonstrator. The computer, power boards, and the digitizer system's Concentration Board (CoBo) are not visible.\label{fig:ReadoutTrinity}}
\end{figure}

\begin{figure}[!htb]
\includegraphics*[angle=0,width=.9\columnwidth]{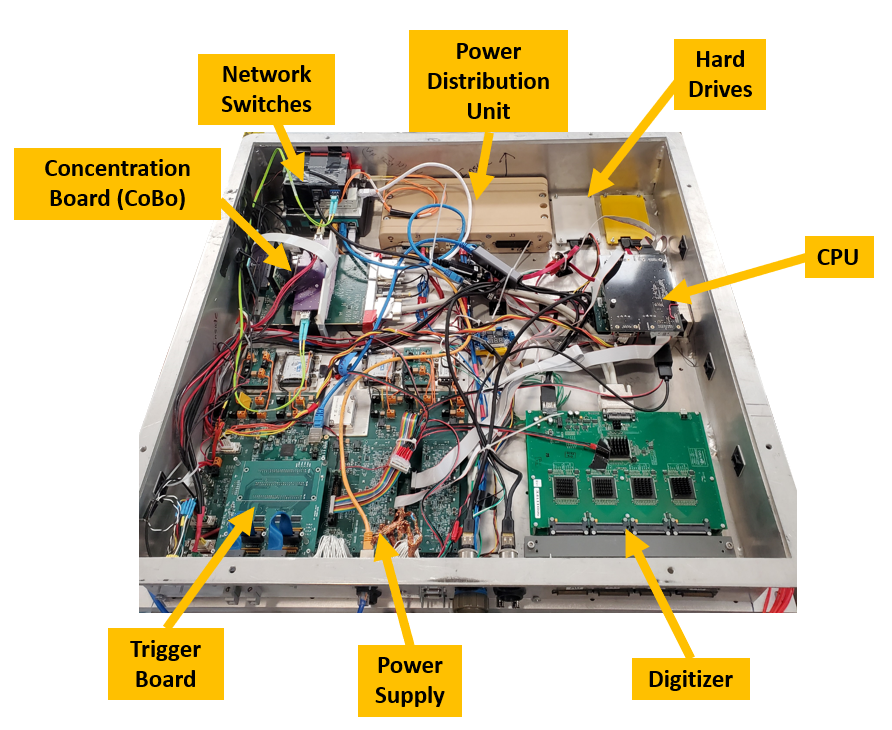}
\caption{The assembled readout unit for EUSO-SPB2.\label{fig:ReadoutSPB2}}
\end{figure}

\paragraph{The readout unit} Figure \ref{fig:CameraModules} shows the block diagram of the readout unit, and Figures \ref{fig:ReadoutTrinity} and \ref{fig:ReadoutSPB2} show pictures of the assembled units for both telescopes. The centerpiece of the readout unit is the central processing unit, also called the camera computer, which serves as the communication gateway to all system components and stores the digitized camera signals. The camera computer interfaces with a high-speed ethernet link to the digitizer system, and a separate network connection is provided to the trigger board. The camera and power supplies are configured via I2C and SPI interfaces.

The readout unit also includes the digitizers for the camera signals, the power supply units, and the trigger board, which we will discuss in more detail in the following sections.

\section{Description of the Main Camera and Readout Components \label{sec:components}}

In this section, we detail the main components of the camera unit and the readout unit: the photosensors, the SIAB front-end electronic boards, the camera computer, the digitizer, the trigger board, and the power supply boards.  

\subsection{Photosensors}

\begin{figure}[!htb]
    \begin{center}
    \includegraphics[width=.8\columnwidth]{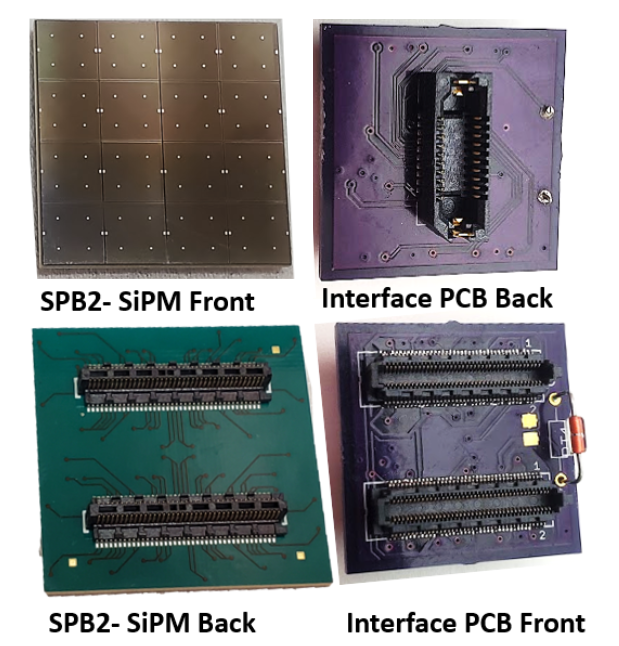}
    \end{center}
    \caption{The front and back side of the S14521-6050AN SiPM matrix used in the EUSO-SPB2 camera. The front and back of the Connector Adaptor populated with a thermistor is shown on the right.
    \label{fig:S14521}}
\end{figure}

\begin{figure}[!htb]
    \begin{center}
    \includegraphics[width=.8\columnwidth]{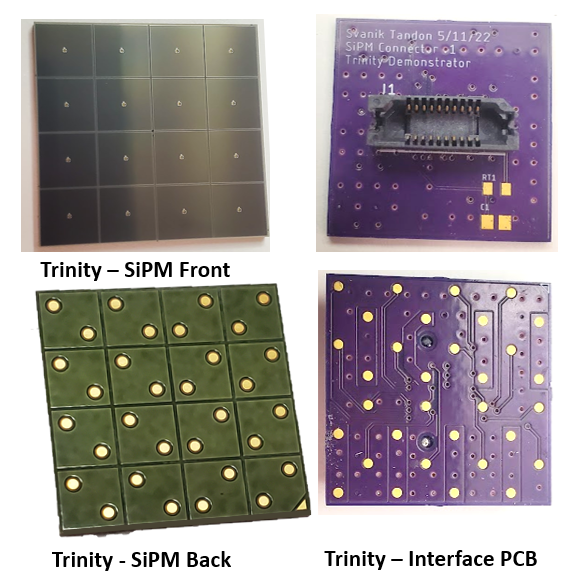}
    \end{center}
    \caption{The front and back side of the S14161-6050HS SiPM matrix used in the \emph{Trinity} Demonstrator. The front and back of the Interface PCB onto which we soldered the SiPM matrix is shown on the right. \label{fig:S14161}}
\end{figure}

Imaging air showers with Cherenkov light is best done with fast, single-photon resolving photosensors to improve the separability of air-shower signals from NSB fluctuations. Natural photosensor candidates are bialkali photomultiplier tubes (PMTs) or silicon photomultipliers (SiPMs). We chose SiPMs because their spectral response is a better fit for the red-peaking Cherenkov spectrum of air showers developing in the lower atmosphere a hundred kilometers from the \emph{Trinity} telescopes \cite{Otte2019d}.

After evaluating devices from different vendors, we chose the Hamamatsu S14520-6050CN, which has high efficiency in the red, low 1.5\% optical crosstalk, low afterpulsing, and only $\sim 0.5\%/^\circ$C gain drift. The SiPMs are 6.4\,mm$\times$6.4\,mm in size and composed of $50\,\mu$m cells. The SiPMs came assembled in $4\times4$ matrices closely packed to minimize the dead space between SiPMs and with minimal dead space at the matrices' edges. Measurements of the S14520-6050CN we did during the selection process are presented in section \ref{sec:SiPMCharacterization}. The actual SiPMs integrated into the cameras are minor evolutions of the S14520-6050CN, the S14521-6050AN is integrated into the EUSO-SPB2 camera, and the S14161-6050HS is integrated into the \emph{Trinity} Demonstrator. The basic characteristics of both series are comparable to the S14520-6050CN. 

Figure \ref{fig:S14521} shows pictures of the front and back of a S14521-6050AN matrix. To interface with the matrix, which has two connectors on its back, we designed the Connector Adaptor printed circuit board (PCB) shown on the figure's right side. Also visible in the picture of the Connector Adaptor is a wired thermistor, which pushes against the SiPM matrix when the matrix is plugged into the Connector Adaptor. Similarly, Figure \ref{fig:S14161} shows pictures of the front and back of a S14161-6050HS matrix used in the \emph{Trinity} Demonstrator and the Interface PCB onto which we reflow soldered the matrices because they came without connectors. The interface PCB also has a surface mounted thermistor, which is not yet populated onto the PCB shown in the picture.

\subsection{Sensor Interface and Amplification Board}

\begin{figure}[!htb]
    \begin{center}
    \includegraphics[width=.8\columnwidth]{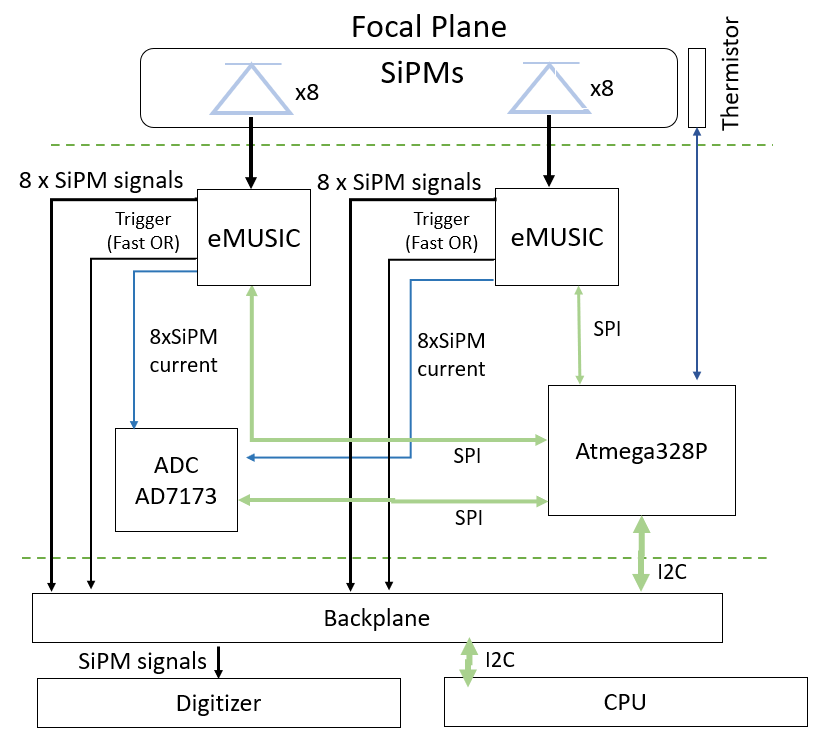}    
    \end{center}
    \caption 
    { \label{fig:SIAB_diagram} Block Diagram of the SIAB}
    \end{figure}
    
\begin{figure}[!htb]
    \begin{center}
    \includegraphics[width=\columnwidth]{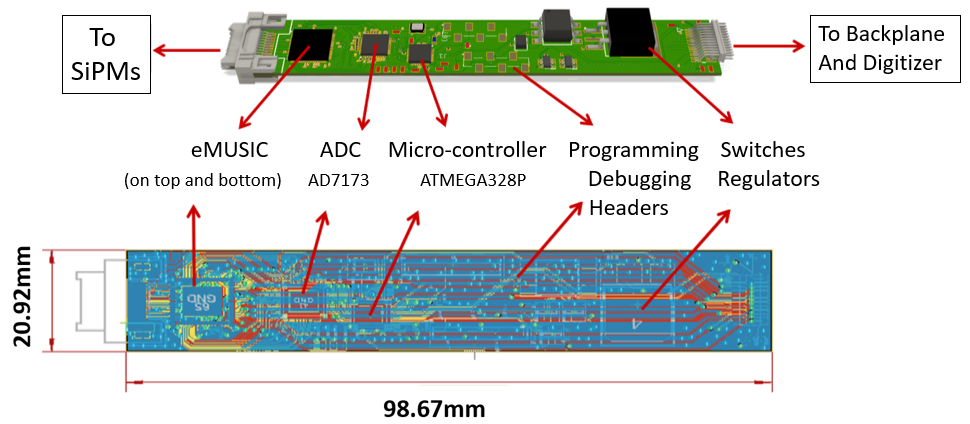}  
    \end{center}
    \caption 
    { \label{fig:SIAB_CAD} PCB of the SIAB }
\end{figure}

The SiPM matrices connect with their adaptor boards and, in the case of EUSO-SPB2, with the additional angle adaptor PCB (see Figure \ref{fig:SPB2CAD}) to the Sensor Interface and Amplification Boards (SIABs). The SIABs amplify and shape the SiPM signals with the Multipurpose Integrated Circuit (eMUSIC) Application Specific Integrated Circuits (ASICs), which is designed as the front end for SiPMs in Cherenkov telescope applications \cite{Gomez2016}.
Figure \ref{fig:SIAB_diagram} shows the block diagram of the SIAB, and Figure \ref{fig:SIAB_CAD} shows the CAD drawing and the electrical layout of the SIAB with signal lines shown in blue.

Each SIAB has an eMUSIC on the top of its PCB and one on the bottom side because the eMUSIC has only 8 input channels, but a SiPM matrix has 16 pixels. Besides shaping and amplifying the SiPM signals, the eMUSIC also has a leading-edge discriminator for each channel, which we use to derive the trigger for the readout, as we explain in Section \ref{sec:trigger}. The signals of all discriminators are OR'd, and only the OR'd signal is available as an output of the eMUSIC. The eMUSIC, furthermore, provides a SiPM bias trim voltage for each channel that is adjustable over 890\,mV in steps of 3.2\,mV, and it features a current monitor output for each channel which we digitize with an AD7173 16-channel analog-to-digital converter on the SIAB.

The eMUSICs and the AD7173 are configured and monitored via SPI by an Atmega328p microcontroller. (see Figure \ref{fig:SIAB_diagram}). The Atmega328p also records the thermistor values, it controls the 3.3\,V and 5\,V regulators that bias the eMUSICs and the ADC, and it turns the SiPM bias voltage on and off. The microcontroller is connected to an I2C bus to communicate with the camera computer. 

We placed the eMUSICs as close as possible to the SiPM-facing end of the SIAB PCB to minimize the pick up of electronic noise. We also wanted the SiPM matrices to directly connect to the SIABs, which constrained the width of the SIAB boards to be less than the size of a SiPM matrix (see Figure \ref{fig:SIAB_CAD}). The SIAB width is 21\,mm about 4.6\,mm narrower than the size of a SiPM matrix.

\subsection{Digitizer}

The SiPM signals connect via 2\,m long micro-coaxial cables from Samtec into an ASIC for General Electronics for TPC (AGET) based digitizer system \cite{Pollacco2018}. The AGET is a 64-channel switched capacitor array (SCA) ASIC with a buffer depth of 512 cells that is sampled with 100\,MS/s and, therefore, records 5.12\,$\mu$s long traces. When the digitizer system receives a readout command from the trigger, the analog signals in the SCA are digitized with 12-bit resolution and transferred into the camera computer. Important for EUSO-SPB2 was the low power consumption of the AGET system of $<10$\,mW per channel.

The AGET ASICs are integrated into groups of four on ASIC Support \& Analog-Digital conversion (AsAd) boards, providing 256 channels per board. Up to four AsAd boards are connected to a Concentration Board (CoBo), which, besides managing the AsAd boards and collecting the digitized traces, applies time stamps, zero suppression, and compression algorithms to the digitized signals. We use one AsAd board for the \emph{Trinity} Demonstrator and two boards for EUSO-SPB2.

The AGET is designed to digitize signals from time projection chambers, which produce much slower signals with rise times on the order of 100\,ns. However, our camera signals have $<10$\,ns rise times, which result in a non-linear response of the AGET. We regained a linear response after inserting a third-order low-pass Butterworth filter with a cut-off frequency of 15\,MHz at the input of each AGET channel, resulting in a 20\,ns rise time.

\subsection{Trigger\label{sec:trigger}}

The command to digitize and save an event comes from the trigger system, which continuously searches the camera signals for signatures that could be due to an air-shower-generated Cherenkov-light flash. We implemented a two-level trigger, where the first level is the leading-edge discriminators in the eMUSICs. The signals from the discriminators in each eMUSIC are combined in a logic OR and provided as one output signal of the eMUSIC. With only the OR'd signal, the trigger system does not know which of the eight pixels connected to the eMUSIC has a signal above the discriminator threshold.

\begin{figure}[!htb]
    \begin{center}
    \includegraphics[width=\columnwidth]{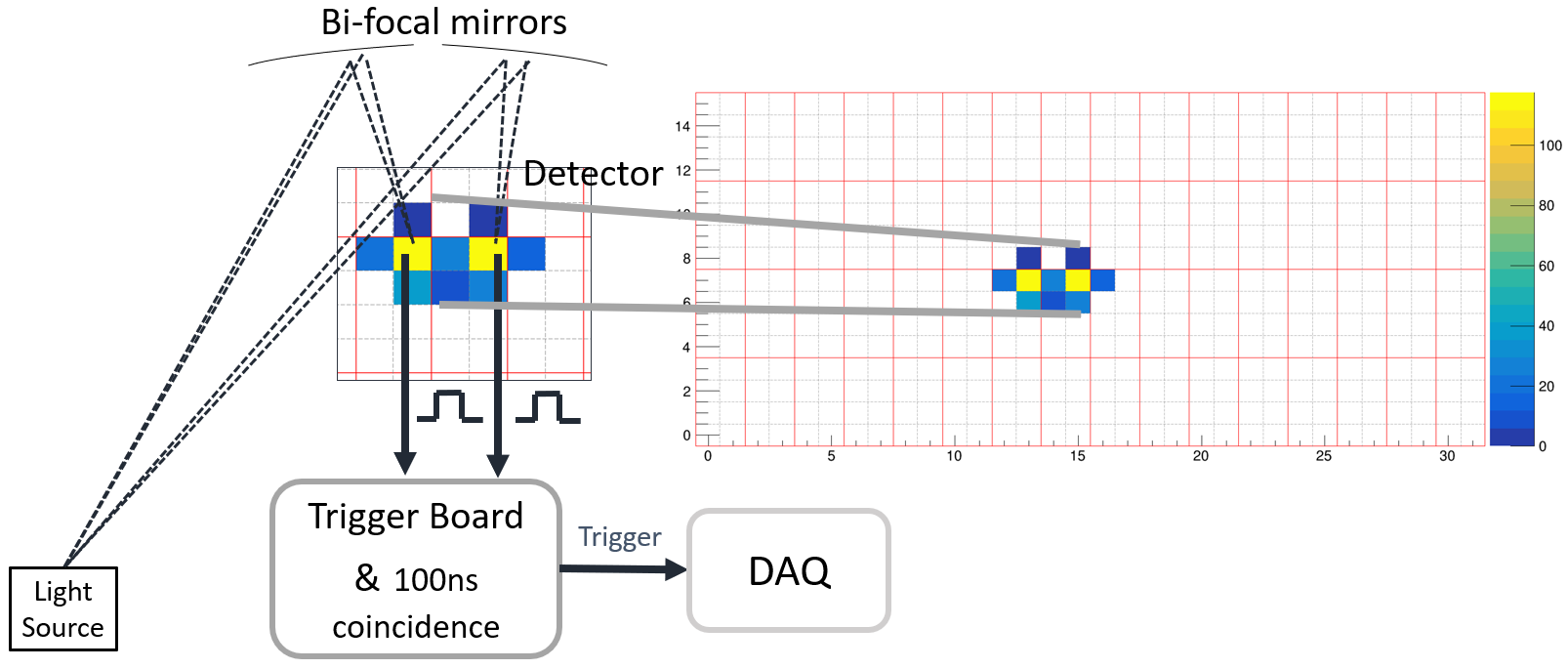}  
    \end{center}
    \caption 
    {\label{fig:TriggerTopology} Illustration of the trigger topology implemented in the trigger for EUSO-SPB2. The pixels within one red rectangle are connected to the same eMUSIC.}
\end{figure}

In the EUSO-SPB2 camera, we have mapped the camera pixels to eMUSIC inputs as indicated by the red rectangles in Figure \ref{fig:TriggerTopology}. The EUSO-SPB2 optics are split to form bi-focal optics, which means that the four mirror segments of the primary mirror are aligned such that an object is imaged twice on the camera with a separation of two pixels. To illustrate the split optics, the figure gives an example of a point source at infinity that gets imaged twice onto the camera. Because of how we have mapped the camera pixels into the eMUSICs and because the image copies are separated by two pixels, there will always be two discriminators in two neighboring eMUSICs that trigger. We exploit this spatial and temporal coincidence in EUSO-SPB2 by requiring a temporal 100\,ns coincidence between the eMUSIC outputs of two neighboring eMUSICs to trigger the readout. The bi-focal optics help us better reject events due to fluctuation in the NSB.

We do not split the optics in the \emph{Trinity} Demonstrator because we are much closer to the air shower and, therefore, can spatially resolve it with the Demonstrator's $0.3^\circ$ pixel size \cite{Otte2019d}. Because the eMUSIC only provides the OR'd discriminator output of 8 channels, we could not implement a standard next neighbor trigger commonly used in VHE gamma-ray Cherenkov telescopes \cite{Zitzer2013,Paoletti2007,Funk2004}. Without an additional coincidence requirement, we, thus, record an event whenever a discriminator produces an output signal. However, we require a spatially extended air-shower image in the event reconstruction or otherwise reject the event.

The discriminator eMUSIC outputs are connected to a Mesa Electronics' 7I80HD-25 Field Programmable Gate Array (FPGA) Ethernet Anything I/O card. The 7I80HD-25 has 72 programmable I/O pins, of which we configured 64 inputs for EUSO-SPB2 and 32 inputs for the \emph{Trinity} Demonstrator, respectively, one for each eMUSIC. Programmed into the FPGA is a finite state machine, which constantly evaluates the input signals and sends a logic signal to the CoBo, triggering the readout of the digitizer whenever the trigger condition is met.  In the split-optics case of EUSO-SPB2, the trigger condition is a 100\,ns time coincidence between the signals of two neighboring eMUSICs. In \emph{Trinity}'s case, the FPGA just passes the eMUSIC logic signals through, triggering the readout whenever it records a discriminator signal without requiring an additional coincidence.

In addition, the trigger board can simultaneously trigger an external LED-based light flasher and the readout system. The flasher provides signals to calibrate and monitor the camera's health.

\subsection{Power Supply and Distribution}
Power for the camera electronics and the SiPMs is provided by a custom dual power system consisting of the Low-Voltage Power Supply (LVPS) and the High-Voltage Power Supply (HVPS) boards. Power for the LVPS/HVPS boards comes from an external 18-30\,V source, which is an array of batteries in the case of EUSO-SPB2 and a programmable SL32-46/U power supply from MAGNA Power in the case of the \emph{Trinity} Demonstrator.

The LVPS board integrates multiple DC-DC converters that generate the +3.3V, +5V, and +12V needed for the different components of the camera and the readout. In EUSO-SPB2, the LVPS also powers the camera computer and the CoBo module of the AGET digitizer. The voltages and currents on the LVPS board are monitored with four 24-bit 16-channel Analog-to-Digital (ADC) converters (AD7173-8BCPZ).

The HVPS board generates the SiPM bias voltages in eight independent high-voltage (HV) channels, each channel powering 64 SiPMs. The voltage of each channel is adjustable with 16-bit Digital-to-Analog (DAC) converters (AD5686R) up to 50\,V, with a resolution of 1\,mV. The current of each HV channel is limited to 20\,mA to protect the eMUSICs, which are damaged if the SiPM currents are too high.

The EUSO-SPB2 system also included a power distribution unit, which was commanded by the camera computer via CAN bus and controlled the power to several auxiliary systems.

\subsection{Camera Computer}
The camera computer for EUSO-SPB2 was a dual-core single-board computer from RTD Embedded Technologies (CMA24CRD1700HR). The camera computer for the \emph{Trinity} Demonstrator is a Sintrones EBOX-7000 edge computing device designed for industrial automation that is passively cooled and operates over a wide temperature range of $-40^\circ$C - $70^\circ$C. Its CPU is an Intel Gen9 Core i7-9700TE (12MB Cache 1.8GHz up to 3.8GHz). It has 8GB of RAM and a 10-minute backup UPS. The EBOX includes two PCIe 3.0 x 8 slots, with one slot currently housing a 10 GB SFP network card.

The camera computers receive the digitized waveforms from the AGET system and perform several management and monitoring tasks. They send commands to the Trigger board and the AGET digitizer via an Ethernet connection and to the SIAB microcontrollers and LVPS/HVPS boards via a System Management Bus (SMBus or I2C). 

The camera computer's control software is programmed in C++ and relays commands to the corresponding processes running on the camera computer using POSIX message queues. Replies are sent back in a separate message queue. 

\section{Cooling System and Thermal Vacuum Testing\label{sec:cooling}}

\begin{figure}[!htb]
    \begin{center}
    \includegraphics[width=\columnwidth]{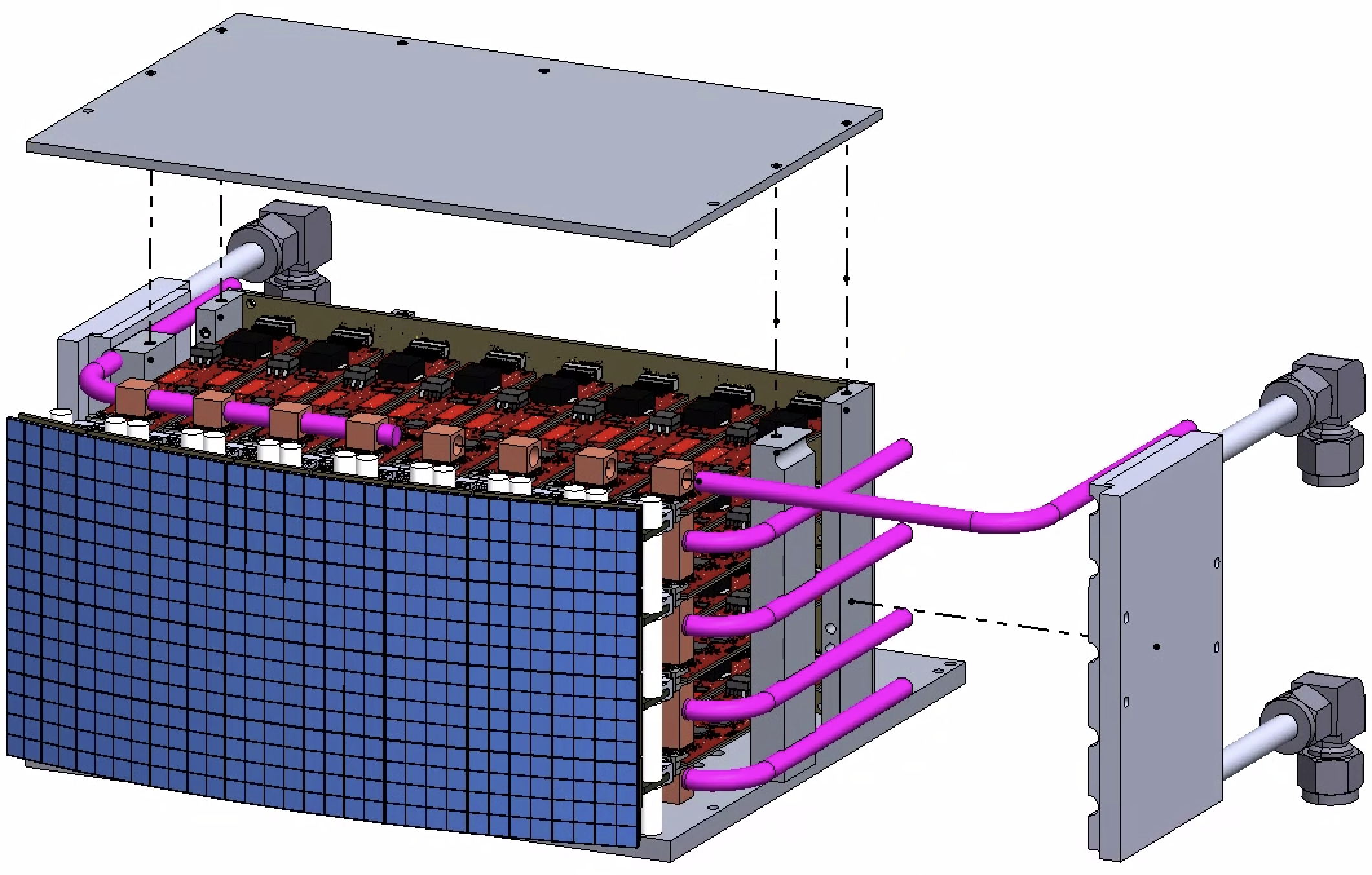}  
    \end{center}
    \caption 
    { \label{fig:CameraCoolingConcept} Exploded view of the EUSO-SPB2 camera showing the heat pipes (purple), the copper blocks (brown) coupled to the eMUSICs, and the cold plate (grey) the heat pipes couple to.}
\end{figure}

\begin{figure}[!htb]
    \begin{center}
    \includegraphics[width=\columnwidth]{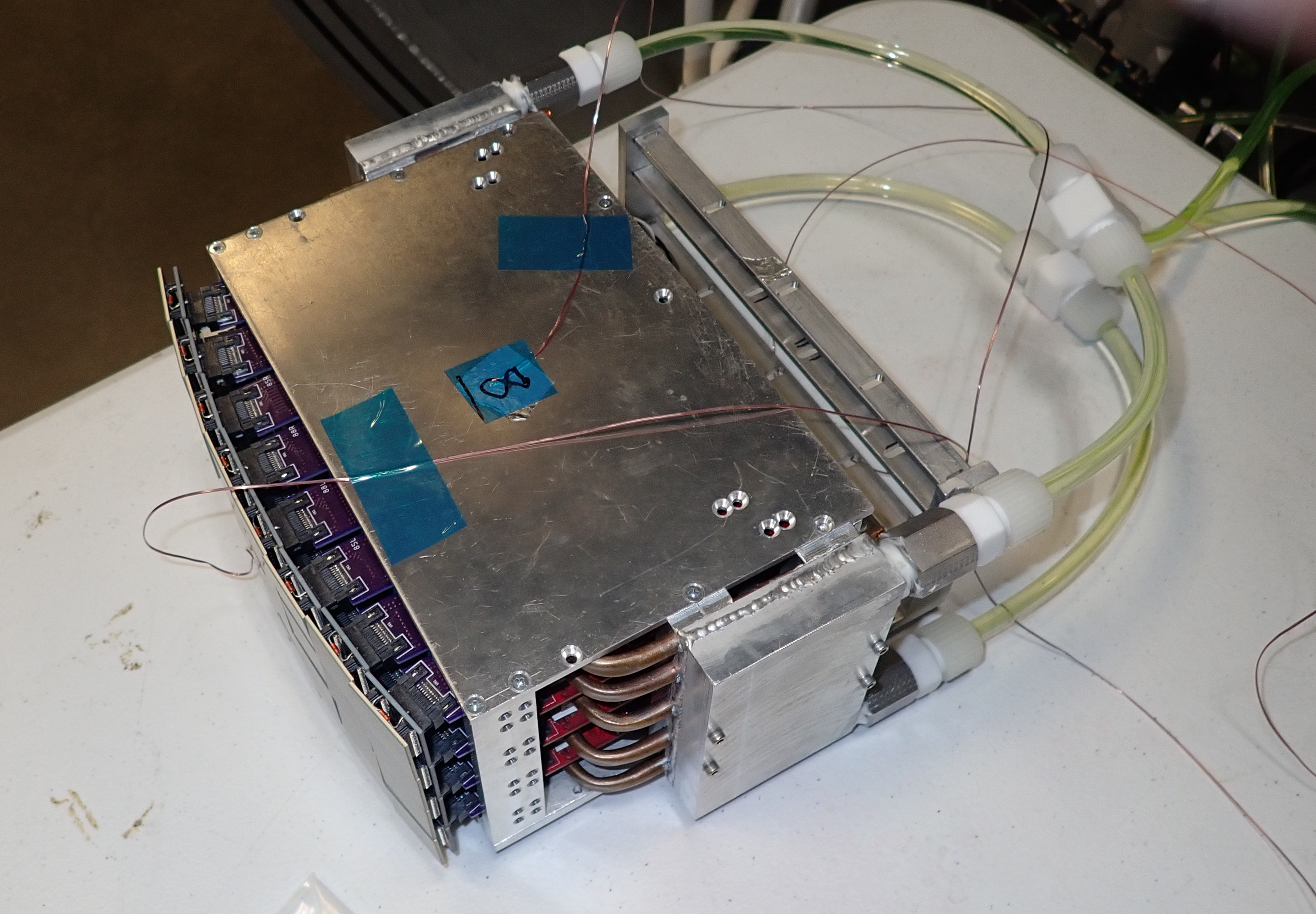} 
    \end{center}
    \caption 
    { \label{fig:coolingSystem} Fully assembled EUSO-SPB2 camera. The heat pipes coming out on the sides are coupled to liquid-cooled cold plates. The tubing filled with glycol antifreeze connects to the cold plates. The wires are probes to monitor the temperatures at different points in the camera during thermal vacuum testing.}
\end{figure}

The design of the cooling system is driven by the requirement to operate the EUSO-SPB2 camera at 33\ km altitude where the ambient pressure is $\sim1$\,hPA, and convective cooling becomes inefficient. We devised a system of heat pipes for the camera unit that transports heat from inside the camera to its sides (see Figure \ref{fig:CameraCoolingConcept}). Inside the camera, the heat pipes are thermally coupled with copper blocks to the eMUSIC packages. Each eMUSIC generates 0.50\,W of heat, which, when the power of all eMUSICs is totaled, amounts to about 90\% of the power of the camera unit. On both sides of the camera, the heat pipes couple to liquid-cooled cold plates (see also Figure \ref{fig:coolingSystem}).

\begin{figure}[!htb]
    \begin{center}
    \includegraphics[width=0.9\columnwidth]{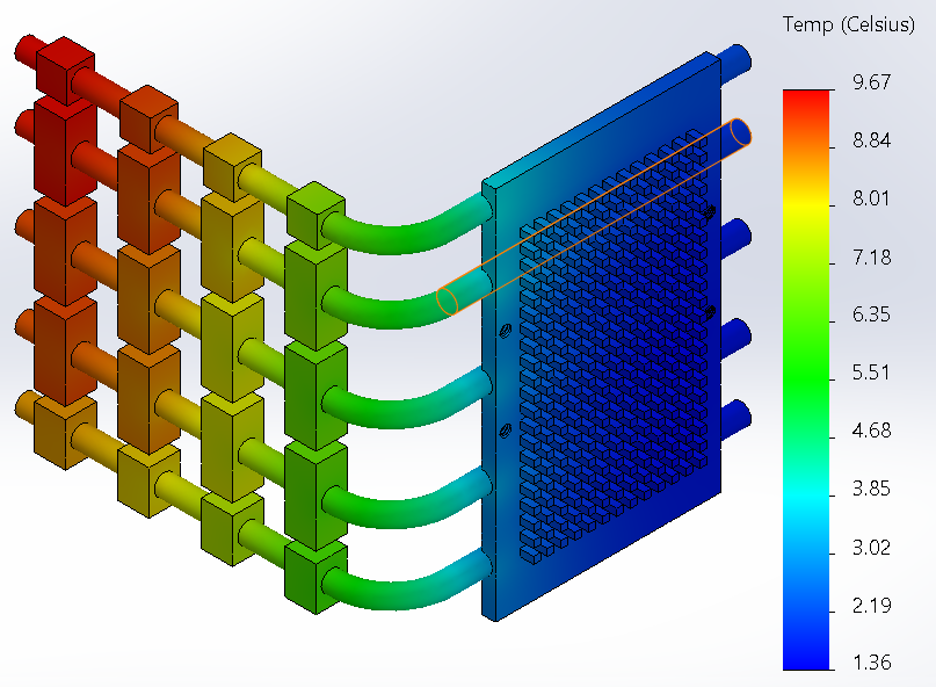}  
    \end{center}
    \caption 
    { \label{fig:CameraCoolingSims}Simulation of the temperature distribution of the camera cooling system. The simulations assume 0.5\,W per eMUSIC, 1\,hPa ambient pressure, and a temperature of the cold plate of $0^\circ$C}
\end{figure}

\begin{figure}[!htb]
    \begin{center}
    \includegraphics[width=0.8\columnwidth]{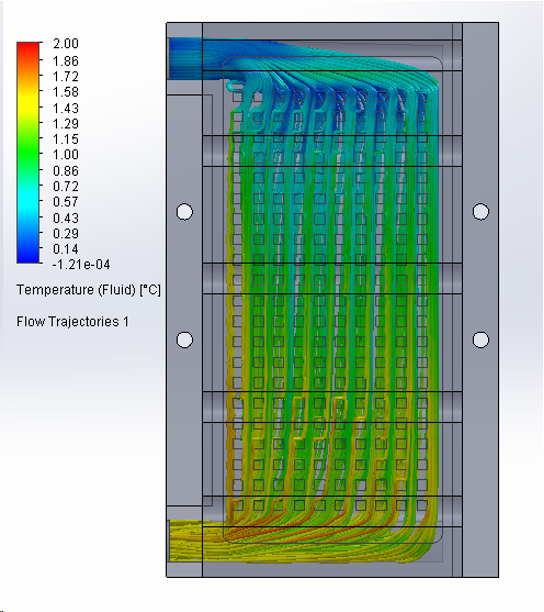} 
    \end{center}
    \caption 
    {\label{fig:ColdPlateSims}Simulation of the fluid flow through the cold plate and the corresponding temperature distribution. The liquid flows from top to bottom at a rate of 0.2\,l/min, and the liquid's temperature is $0^\circ$C at the top.}
\end{figure}

Figure \ref{fig:CameraCoolingSims} shows simulations of the temperature distribution inside the camera cooling system, and Figure \ref{fig:ColdPlateSims} shows the simulation of the temperature distribution of the cooling liquid inside the cold plate.

\begin{figure}[!htb]
    \begin{center}
    \includegraphics[width=\columnwidth]{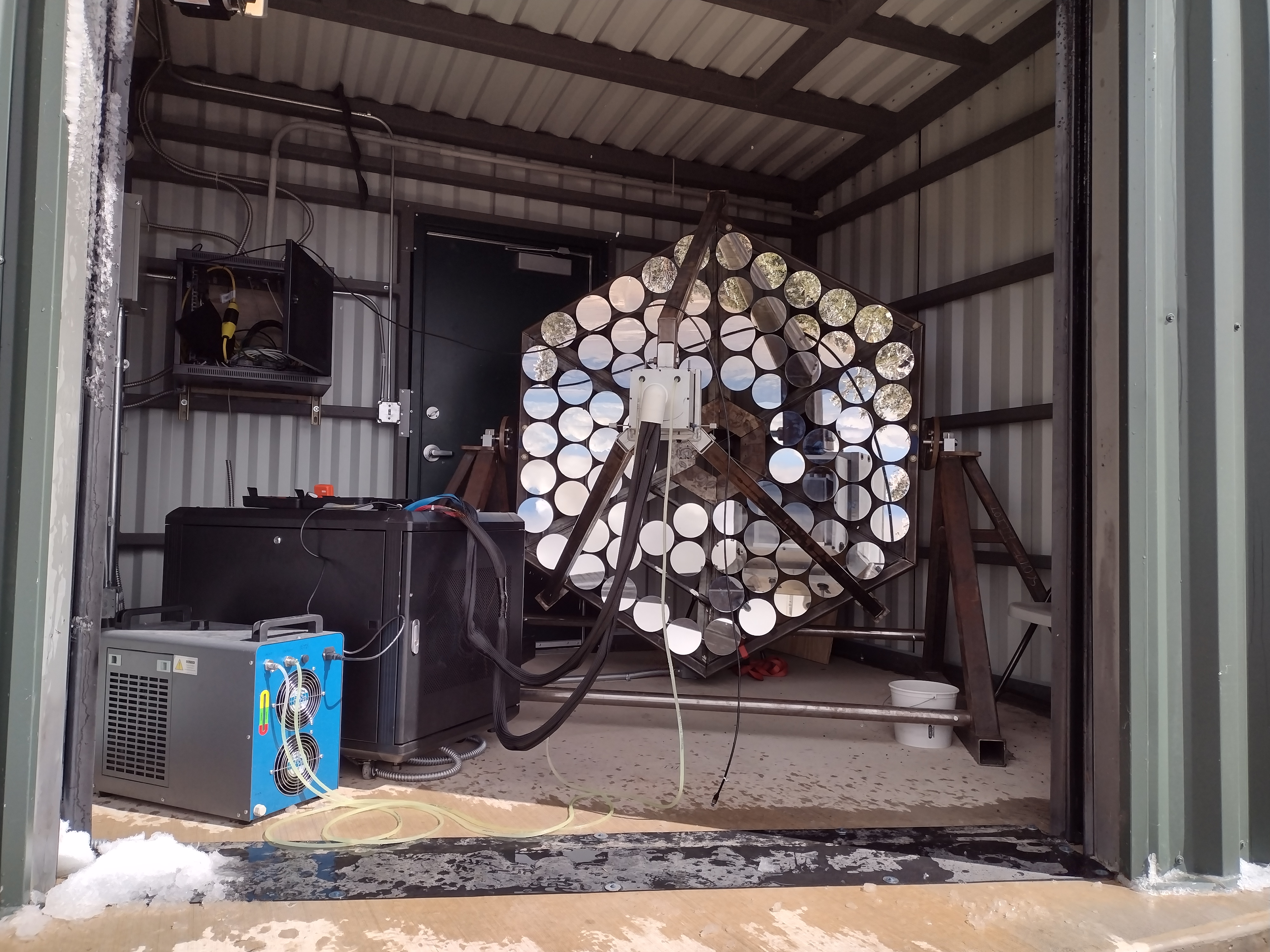} 
    \end{center}
    \caption 
    {\label{fig:Demonstrator} The \emph{Trinity} Demonstrator. In the front left is the Chiller for the cooling of the camera. Behind it is the crate with the readout electronics. The reflection of the camera can be seen in the mirrors of the Demonstrator.}
\end{figure}

\begin{figure}[!htb]
    \begin{center}
    \includegraphics[width=\columnwidth]{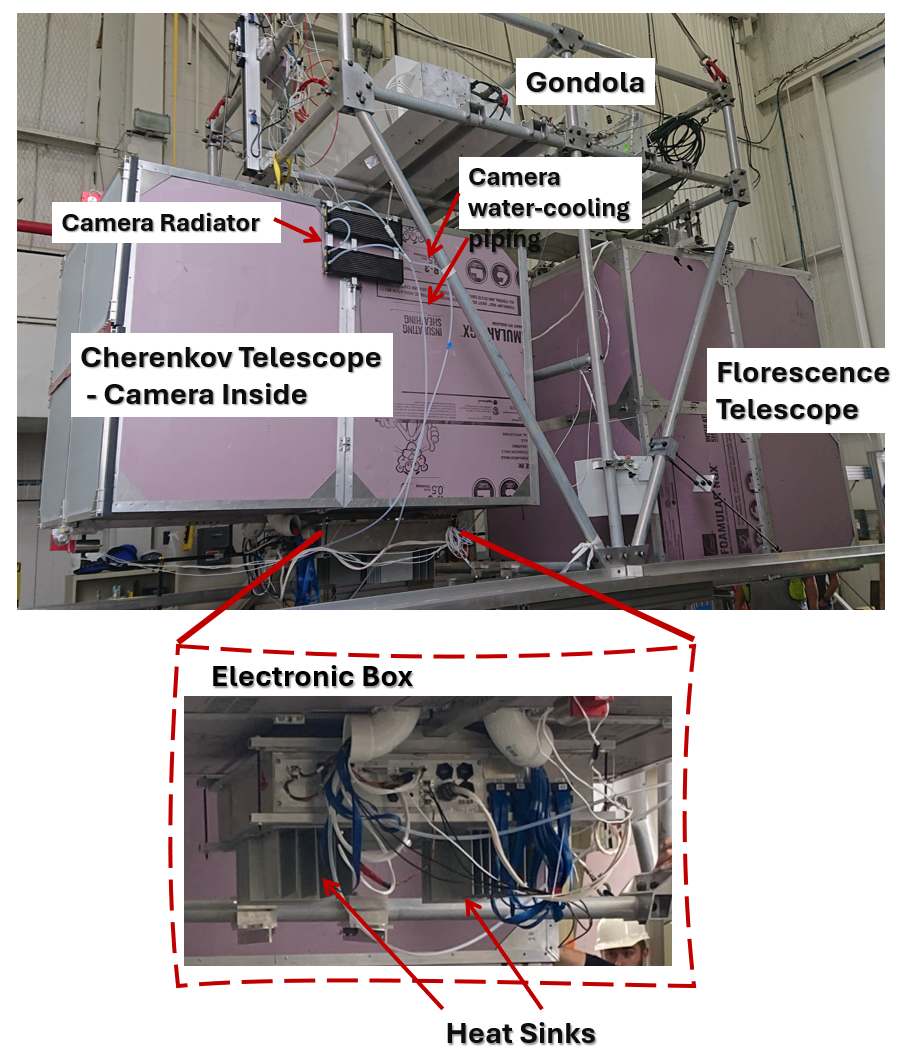} 
    \end{center}
    \caption 
    { \label{fig:MEBHeatSinks} The cooling system of the EUSO-SPB2 camera and readout units. The radiator of the camera cooling system is mounted on the side of the Cherenkov telescope box. The insert shows the readout unit mounted on the bottom of the telescope with its heatsinks facing the Earth.}
\end{figure}

For EUSO-SPB2, the glycol-based antifreeze liquid was circulated by two parallel connected gear pumps ZY-1305 from Speck and circulated through two radiators salvaged from a CPU cooling system, CORSAIR iCUE H150i RGB PRO XT 360\,mm Radiator. The radiators were mounted outside the telescope radiating into space (see Figure \ref{fig:MEBHeatSinks}). In the case of the \emph{Trinity} Demonstrator, the liquid circulates through an OMTech 6L Industrial Water Chiller, set to a temperature of $6^\circ$C (see Figure \ref{fig:Demonstrator}).

\begin{figure}[!htb]
    \begin{center}
    \includegraphics[width=\columnwidth]{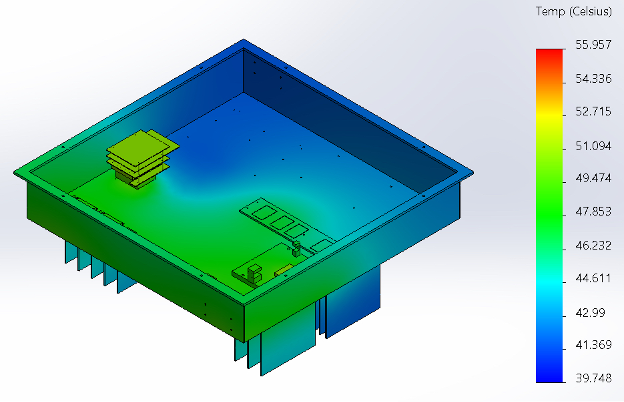}  
    \end{center}
    \caption 
    { \label{fig:MEBSimulations} Simulated temperature distribution inside the enclosure of the readout unit. The simulations assume that the heatsinks face Earth with a surface temperature of $20^\circ$C and an ambient pressure of 1\,hPa.}
\end{figure}

The electronics in the readout unit, i.e., the digitizer, computer, switches, etc, are passively cooled in EUSO-SPB2 with heatsinks mounted on the outside of the enclosure of the readout unit (see Figure \ref{fig:MEBHeatSinks}). For the design of the heatsinks, we thermally modeled the readout unit with all components at their respective positions and their respective power consumptions (see Figure \ref{fig:MEBSimulations}). In the \emph{Trinity} Demonstrator, the digitizer components are cooled by forced air and all other components by convection.

\subsection{Thermal Vacuum Testing}

\begin{figure}[!htb]
    \begin{center}
    \includegraphics[width=0.9\columnwidth]{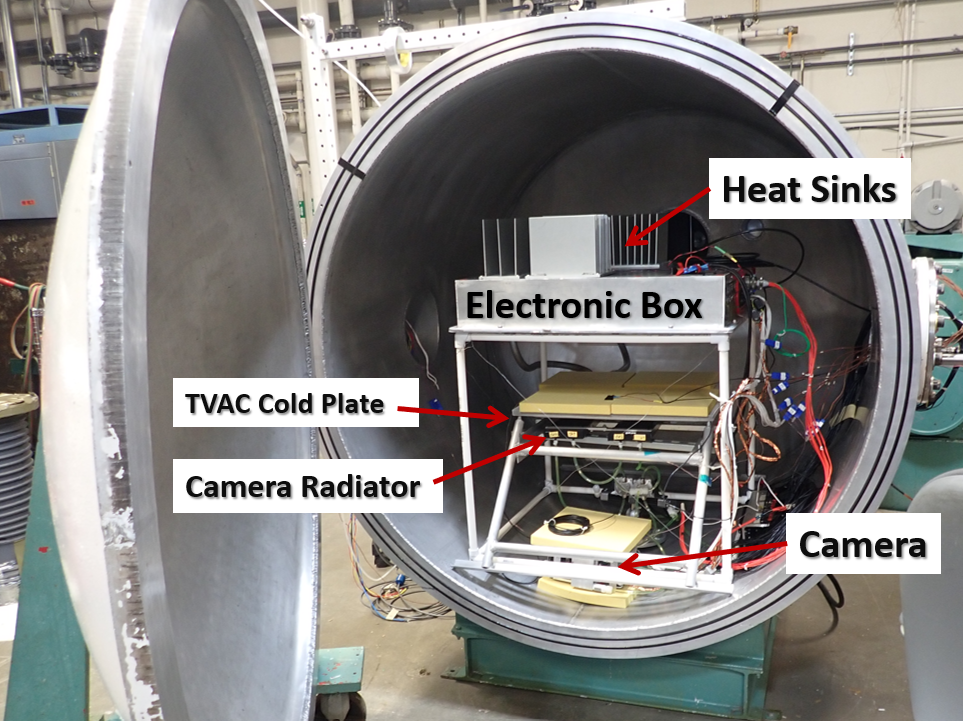} 
    \end{center}
    \caption 
    { \label{fig:TVACTestingSetup}Setup for thermal-vacuum testing the EUSO-SPB2 camera. On top is the electronic box (the readout unit) upside down with the heat sinks facing the top chamber wall. Below the electronic box is the cold plate, whose temperature was varied during testing. The electronic box is thermally shielded from the cold plate with insulation foam (yellow plates). Right below the cold plate are the two radiators. The camera sits on the bottom of the chamber. Insulation foam on top and the bottom of the camera thermally shields the camera from the wall and the cold plate.}
\end{figure}

We thermal-vacuum tested the EUSO-SPB2 camera and readout over a wide temperature range ($-20^\circ$C - $40^\circ$C) at ambient pressure and at 1\,hPa. Figure \ref{fig:TVACTestingSetup} shows the EUSO-SPB2 camera and readout in the thermal vacuum chamber. The radiator faced a cold plate whose temperature we varied over the aforementioned temperature range, while the temperature of the chamber wall remained at room temperature of $\sim20^\circ$C. 

\begin{figure}[!htb]
    \begin{center}
    \includegraphics[width=\columnwidth]{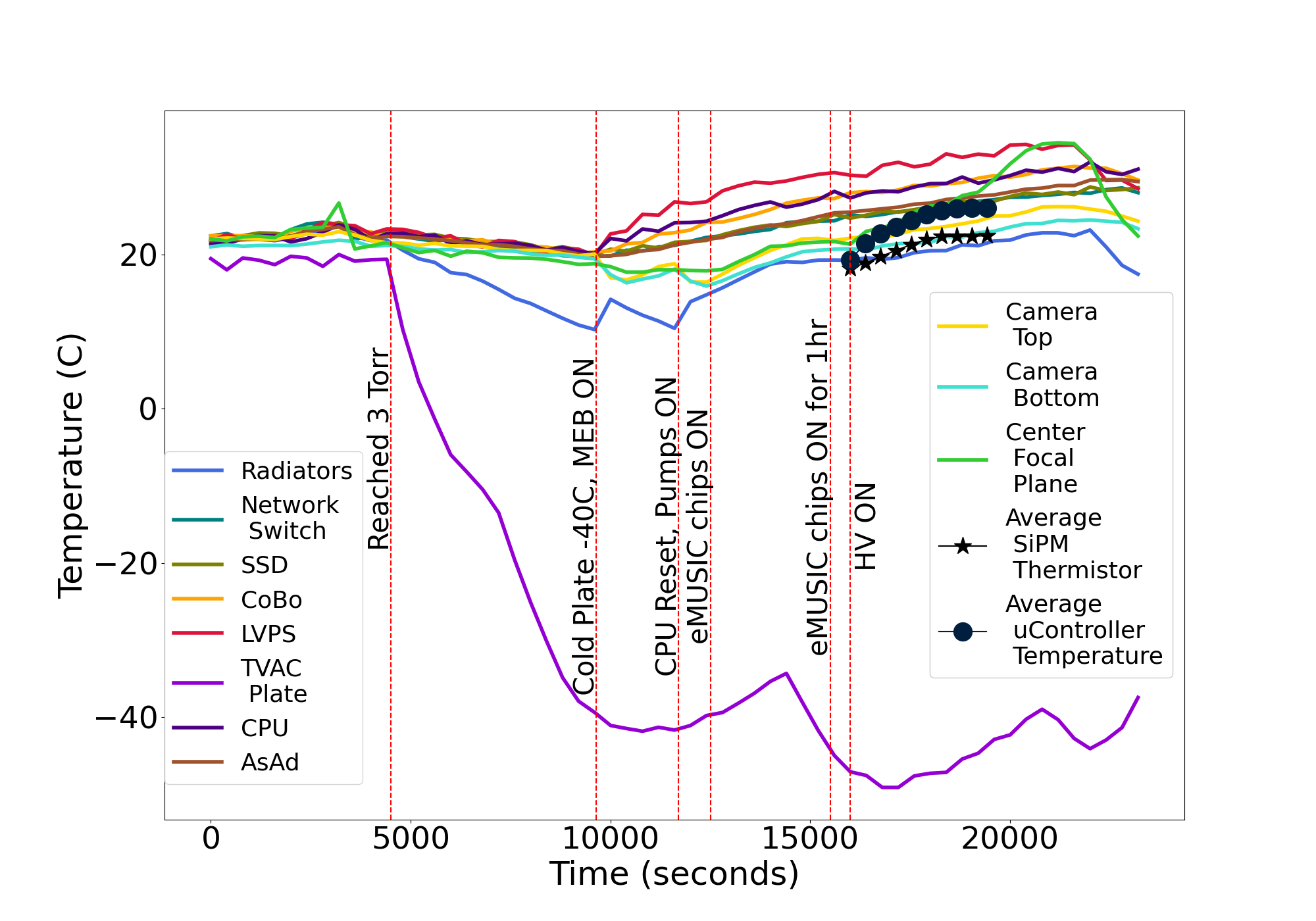} 
    \end{center}
    \caption 
    { \label{fig:TempOrlando}Temperature measurements of the different components of the EUSO-SPB2 camera and readout during thermal vacuum testing.}
\end{figure}

The thermal vacuum tests confirmed our thermal simulations, validated the performance of the EUSO-SPB2 cooling system, and showed that the readout and camera function over the entire tested temperature range at the expected ambient pressure at a float altitude of 33\,km. With the cold-plate temperature at $-40^\circ$C, the SiPM temperature stayed below $30^\circ$ also when we illuminated the SiPMs with a steady light source mimicking expected and extreme night-sky background (NSB) levels, thus keeping the SiPM dark-count rates below the NSB. Figure \ref{fig:TempOrlando} shows temperature measurements of different components of the system during a 6-hour test cycle while the cold plate was held at $-40^\circ$C. 

\section{Photosensor Characteristics and Integration\label{sec:sensors}}

\subsection{Characterization of the S14520-6050CN \label{sec:SiPMCharacterization}}

We evaluated devices from different vendors before purchasing the SiPMs. Here, we present the characteristics of the Hamamatsu S14520-6050CN, which we measured at different temperatures. The measured characteristics include dark count rates, direct and delayed optical crosstalk, effective cell capacitance, quench resistor, recovery time constant, afterpulsing, and breakdown voltage. Based on these measurements, we made the decision to purchase the commercially available versions of the S14520-6050CN, the S14521-6050CN, and the S14161-6050HS. Instead of fully characterizing these, we focused on photon detection efficiency, spectral response, gain, and breakdown voltage measurements. We performed all measurements with the setups and methods described in \cite{Otte2016a}.

\begin{figure}[!htb]
    \begin{center}   
\includegraphics[width=\columnwidth]{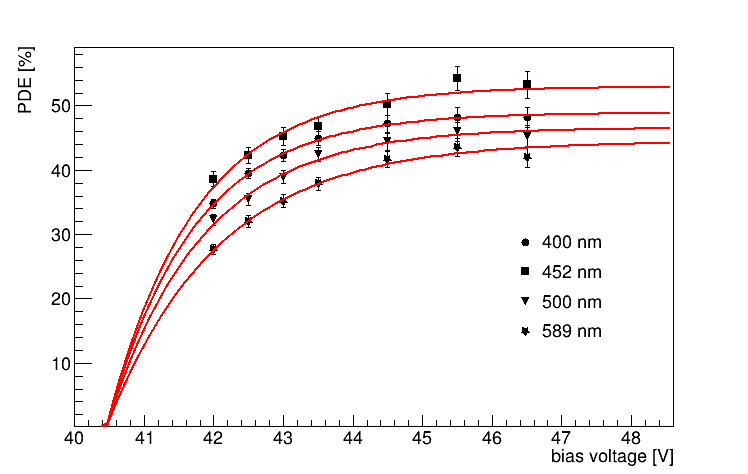} 
    \end{center}
    \caption     { \label{fig:S14520PDE} Photon detection efficiency of the S14520-6050CN vs.\ bias voltage. We operate the SiPMs at the bias where the PDE reaches 90\% of its maximum value. For the S14520-6050CN that bias point at room temperature is at 44\,V or 9\% overvoltage.}
\end{figure}

Figure \ref{fig:S14520PDE} shows the photon detection efficiency (PDE) for four wavelengths as a function of SiPM bias at room temperature. The red lines fit the data points with an exponential function \cite{Otte2018}. From the fit results, we read the bias voltage where the PDE reaches 90\% of the maximum PDE, which is at about 9\% overvoltage for all wavelengths. In the following, we adopt a 9\% relative overvoltage as the operating point of the SiPMs and discuss the impact that the measured SiPM nuisance parameters at that voltage have on the operation of the cameras. The 9\% relative overvoltage operating point is marked with a black arrow on the abscissa in the following figures.

\begin{figure}[!htb]
    \begin{center}   
\includegraphics[width=\columnwidth]{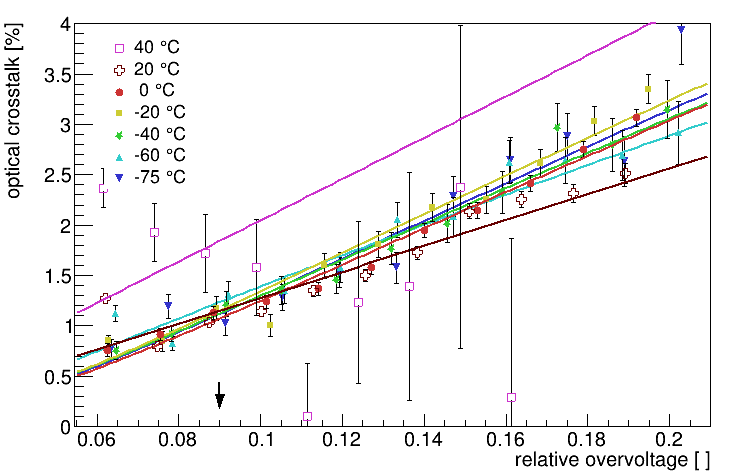} 
    \end{center}
    \caption     { \label{fig:S14520OC} Optical crosstalk vs.\ relative overvoltage of the S14520. The black arrow marks our adopted nominal operating point.}
\end{figure}

Figure \ref{fig:S14520OC} shows the direct optical crosstalk as a function of relative overvoltage. The measurement at $40^\circ$ is unreliable because the high dark count rate at that temperature prevents a clear separation of SiPM signals. A $\sim1$\% optical crosstalk is extremely low and does not impact the camera's performance.

\begin{figure}[!htb]
    \begin{center}   
\includegraphics[width=\columnwidth]{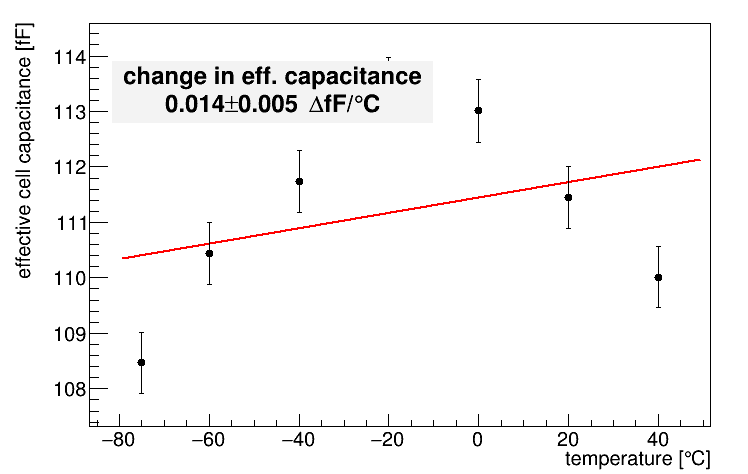} 
    \end{center}
    \caption     { \label{fig:S14520CellCap} Effective cell capacitance vs.\ temperature of the S14520.}
\end{figure}

\begin{figure}[!htb]
    \begin{center}   
\includegraphics[width=\columnwidth]{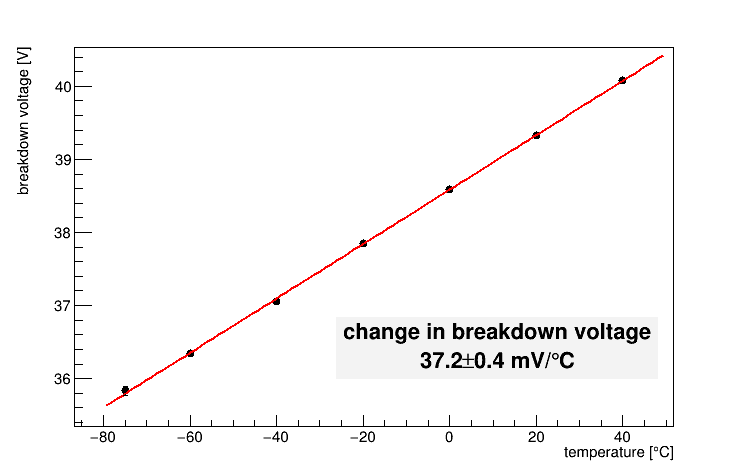} 
    \end{center}
    \caption     { \label{fig:S14520VBD}Breakdown voltage vs.\ temperature of the S14520.}
\end{figure}

Figure \ref{fig:S14520CellCap} shows the effective cell capacitance derived from gain vs.\ bias measurements at different temperatures. The effective cell capacitance is defined as $C=\Delta Q/\Delta U$, i.e.\ gain in charge divided by the overvoltage. The capacitance is stable within $\sim2$\%, which means any gain changes can be solely attributed to the temperature-dependent breakdown voltage, which shows Figure \ref{fig:S14520VBD}.

The breakdown voltage changes linearly with a slope of $37$\,mV/C$^\circ$ between $-75^\circ$C and $40^\circ$. Or normalized to the breakdown voltage with a slope of 0.1\%/C$^\circ$, which is typical for SiPMs. Because we operate the SiPMs with a relative overvoltage of $\sim10$\%, the temperature dependence of the gain at our operating voltage is only 0.1\%/10\%=0.01 per $1^\circ$C or 1\%/$^\circ$C. Such a small temperature dependence significantly reduces the need to keep the temperature of the SiPMs stable or to readjust the SiPM voltages constantly. The same argument also applies to the temperature dependence of the breakdown probability, which factors into the PDE. However, because we operate the SiPMs at a bias where the PDE and thus the breakdown probability are already saturating, the relative changes of the PDE are even less than 1\%/$^\circ$C.

\begin{figure}[!htb]
    \begin{center}   
\includegraphics[width=\columnwidth]{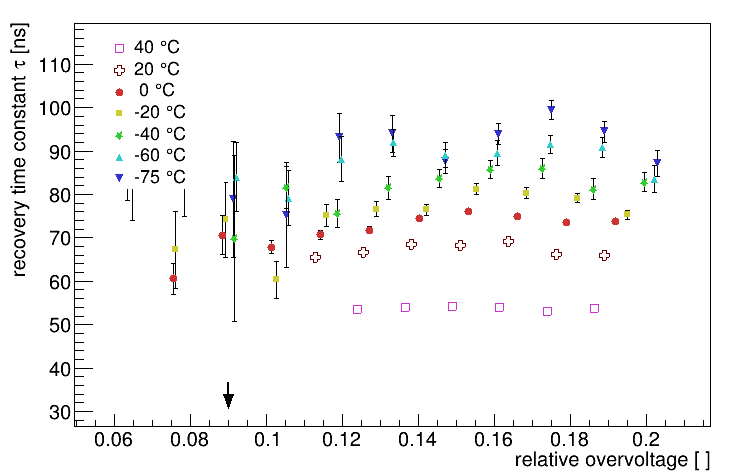} 
    \end{center}
    \caption     { \label{fig:S14520CellRecovery}Cell recovery time constant vs.\ relative overvoltage of the S14520. If error bars are not shown, they are smaller than the symbols.}
\end{figure}

Figure \ref{fig:S14520CellRecovery} shows the charging time constant of a SiPM cell after a breakdown. Within our measurement uncertainties, the time constant does not depend on overvoltage. It does, however, depend on temperature, which we attribute to a changing quenching resistor of $0.3\%/C^\circ$. With a 100\,ns recovery time and an average night-sky background rate of 200\,MHz per SiPM, only 16 cells recover at any time while the remaining $>14,000$ cells of the SiPM can accept photons. The cell recovery time, therefore, does not limit the dynamic range or PDE of the camera.

\begin{figure}[!htb]
    \begin{center}   
\includegraphics[width=\columnwidth]{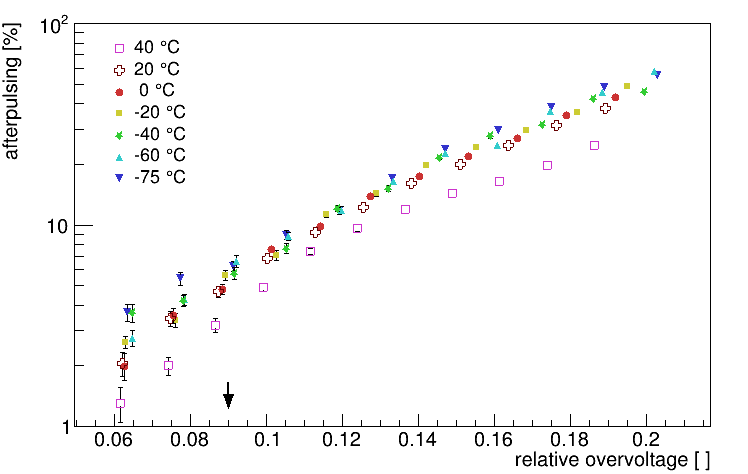} 
    \end{center}
    \caption     { \label{fig:S14520AP}Afterpulsing vs.\ relative overvoltage of the S14520. If error bars are not shown, they are smaller than the symbols.}
\end{figure}

Afterpulsing has a similar impact on the camera performance as direct optical crosstalk. It adds to the primary signal and needs to be accounted for in the event reconstruction. The S14520 (see Figure \ref{fig:S14520AP}) has an acceptable $\sim5\%$ afterpulsing probability at the operating point with a time constant of a few ten nanoseconds.

\begin{figure}[!htb]
    \begin{center}   
\includegraphics[width=\columnwidth]{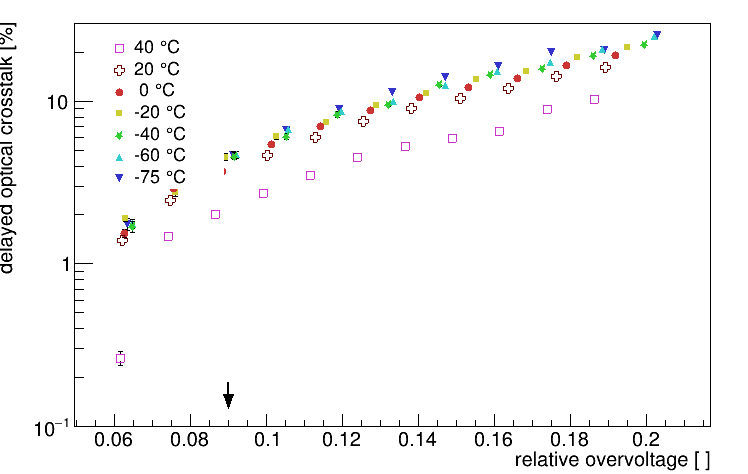} 
    \end{center}
    \caption     { \label{fig:S14520delayedOC}Delayed optical crosstalk vs.\ relative overvoltage of the S14520. If error bars are not shown, they are smaller than the symbols.}
\end{figure}

\begin{figure}[!htb]
    \begin{center}   
\includegraphics[width=\columnwidth]{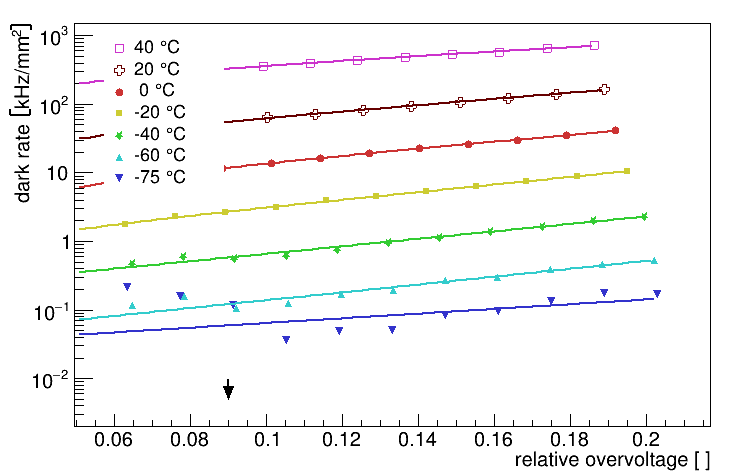} 
    \end{center}
    \caption     { \label{fig:S14520DC}Dark count rates per square millimeter sensor area vs.\ overvoltage of the S14520. The measurement at $-75^\circ$C is affected by systematic effects. If error bars are not shown, they are smaller than the symbols.}
\end{figure}

Delayed optical crosstalk effectively increases dark count rates as it happens significantly after the primary signal. It is less than 5\%, which adds to the dark-count rate by the same amount and thus does not impact the camera performance. That is because the dark-count rates, shown in Figure \ref{fig:S14520DC}, are ten times less than the expected night-sky background rate of 5\,MHz/mm$^2$ even at $40^\circ$C. We expect to operate at $20^\circ$C or below where the dark count rate is $<1$\% of the night-sky background rate. The dark count rates, thus, do not impact the camera's performance.

\begin{figure}[!htb]
    \begin{center}   
\includegraphics[width=0.9\columnwidth]{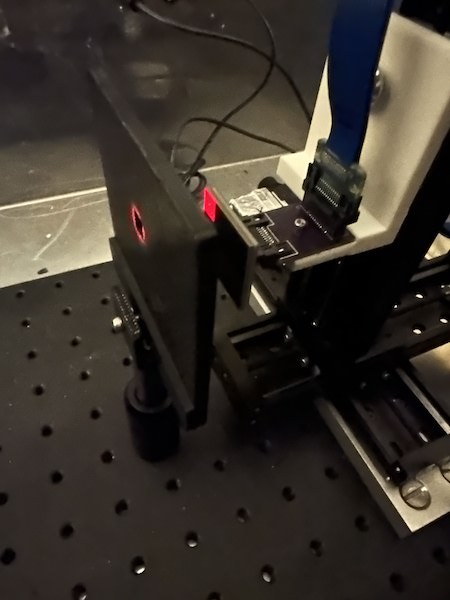} 
    \end{center}
    \caption{\label{fig:SiPMTest}One of the \emph{Trinity} SiPM matrices mounted on a Zaber T-LSM100A x-y stage and illuminated with a picosecond laser at 638\,nm. The laser is shining onto the mask from the left. The illuminated SiPM is seen on the right side of the mask.}
\end{figure}

\subsection{Photon Detector Assembly and Acceptance Testing \label{sec:accepttest}}

After reflow soldering the SiPM matrices for \emph{Trinity} onto their carrier boards, we measured the breakdown voltage and relative gain as part of the acceptance testing. For the testing, the SiPM matrix is mounted onto a T-LSM100A x-y stage from Zaber, as shown in Figure \ref{fig:SiPMTest}, and the stage aligns the SiPM under test with the mask's opening. The size of the opening guarantees that only the SiPM under test is illuminated. The SiPM is flashed with 120\,ps long pulses of 638\,nm light from a Picoquant picosecond laser PDL 800-B with a 638\,nm LDH 8-1-469 laser head. Before the light hits the SiPM, it's intensity is attenuated by a factor of 1,000 with neutral density filters to fit within the linear range of the SiPM. The SiPM signals are amplified with a custom amplifier \cite{Otte2015e} and sampled with 5\,GS/s by a DRS evaluation board \cite{Ritt2010}.

We measured each SiPM's amplitude response at 41\,V and 42\,V bias voltage ($U_{bias}$). Because the amplitude response $A$ of the SiPM is proportional to
\begin{equation}
    A \propto C_{eff}\cdot (U_{bias}-U_{BD}),
\end{equation}
we obtain from these measurements at two bias voltages the average effective cell capacitance $C_{eff}$, and the breakdown voltage $U_{BD}$ for each SiPM.

\begin{figure}[!htb]
    \begin{center}   
\includegraphics[width=\columnwidth]{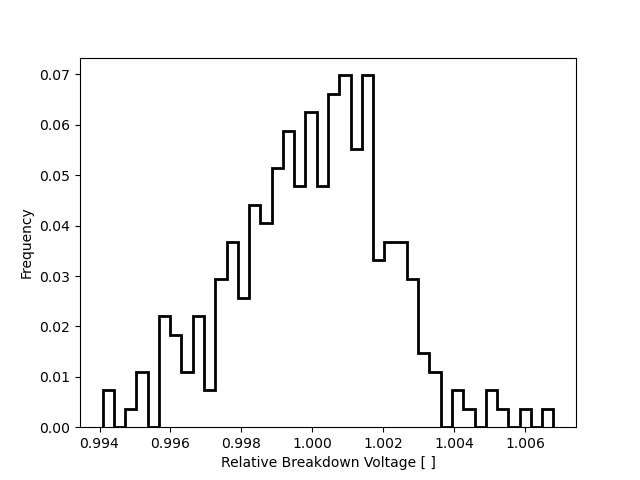}
    \end{center}
    \caption{\label{fig:relBD} Distribution of the breakdown voltages of the \emph{Trinity} Demonstrator SiPMs normalized to the 39.5\,V camera average.}
\end{figure}

Figure \ref{fig:relBD} shows the distribution of the breakdown voltages from these measurements normalized to the 39.5\,V average breakdown voltage of all the \emph{Trinity} SiPMs. The breakdown voltage distribution is sufficiently narrow, so SiPM matrices did not have to be grouped by similar breakdown voltage and all SiPM matrices are biased with the same global HV.

\begin{figure}[!htb]
    \begin{center}   
\includegraphics[width=\columnwidth]{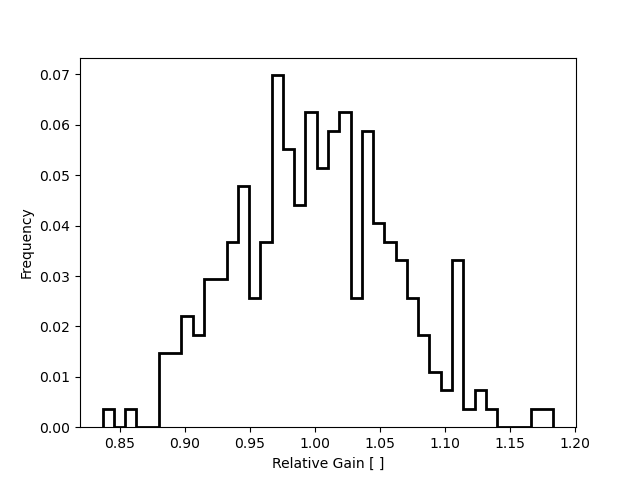} 
    \end{center}
    \caption{\label{fig:relGain} Distribution of the gain of the \emph{Trinity} SiPMs when operated at the same overvoltage. The individual SiPM values are normalized to the camera average.}
\end{figure}

Figure \ref{fig:relGain} shows the relative gain distribution of the SiPMs for the EUSO-SPB2 camera normalized to the camera average. The distribution is due to differences in the effective cell capacitances.

For operating the camera, the bias voltages of the SiPMs are adjusted with the SiPM bias trim voltages provided by the eMUSICs to produce a uniform response of the camera to external light, which we discuss in section \ref{sec:flatfielding}.

\subsection{Photon Detection Efficiency Measurements}

A careful evaluation of the photon detection efficiency (PDE) is necessary because it is a critical factor in determining the telescope's photon collection efficiency. Uncertainties in the PDE worsen the energy resolution and result in systematic offsets in the energy scale during event reconstruction. All PDE measurements have been performed with the setups described in \cite{Otte2016a} and directly feed into the Monte Carlo simulations for the \emph{Trinity} Demonstrator and EUSO-SPB2.

\begin{figure}[!htb]
    \begin{center}   
\includegraphics[width=\columnwidth]{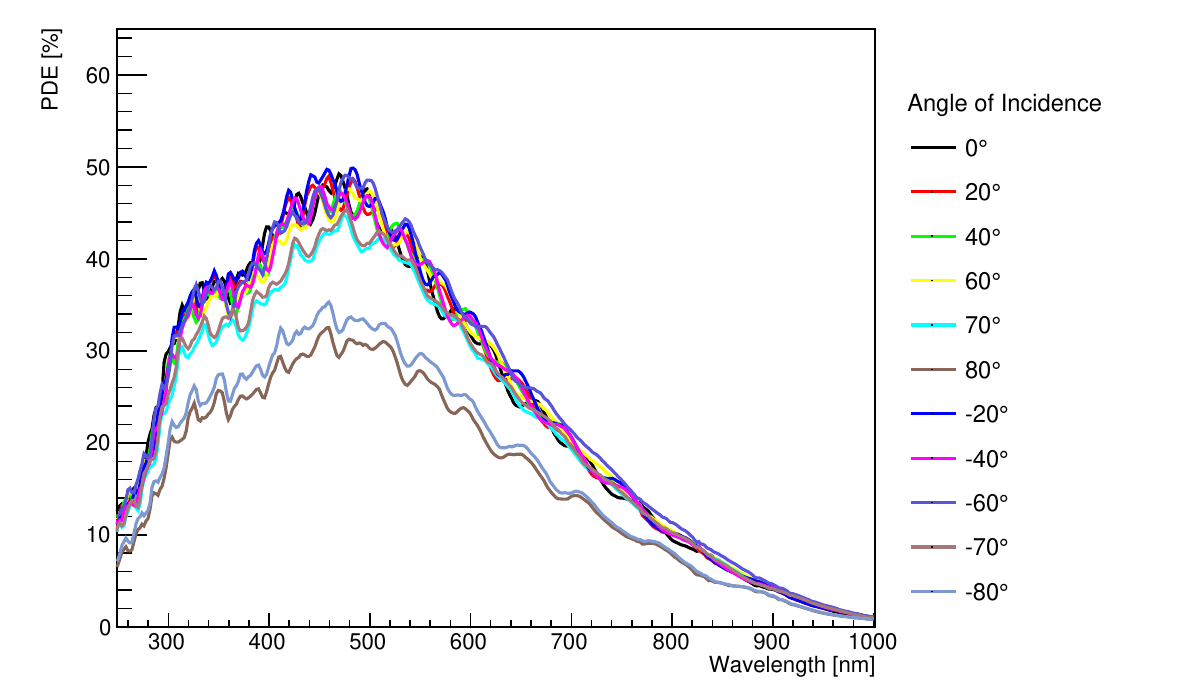} 
    \end{center}
    \caption{\label{fig:PDEvsAngle} Photon Detection Efficiency of the S14521 used in EUSO-SPB2 vs.\ wavelength for normal up to $80^\circ$ angle of incidence. The difference between positive and negative angles is due to small misalignments of the angle of incidence.}
\end{figure}

\begin{figure}[!htb]
    \begin{center}   
\includegraphics[width=\columnwidth]{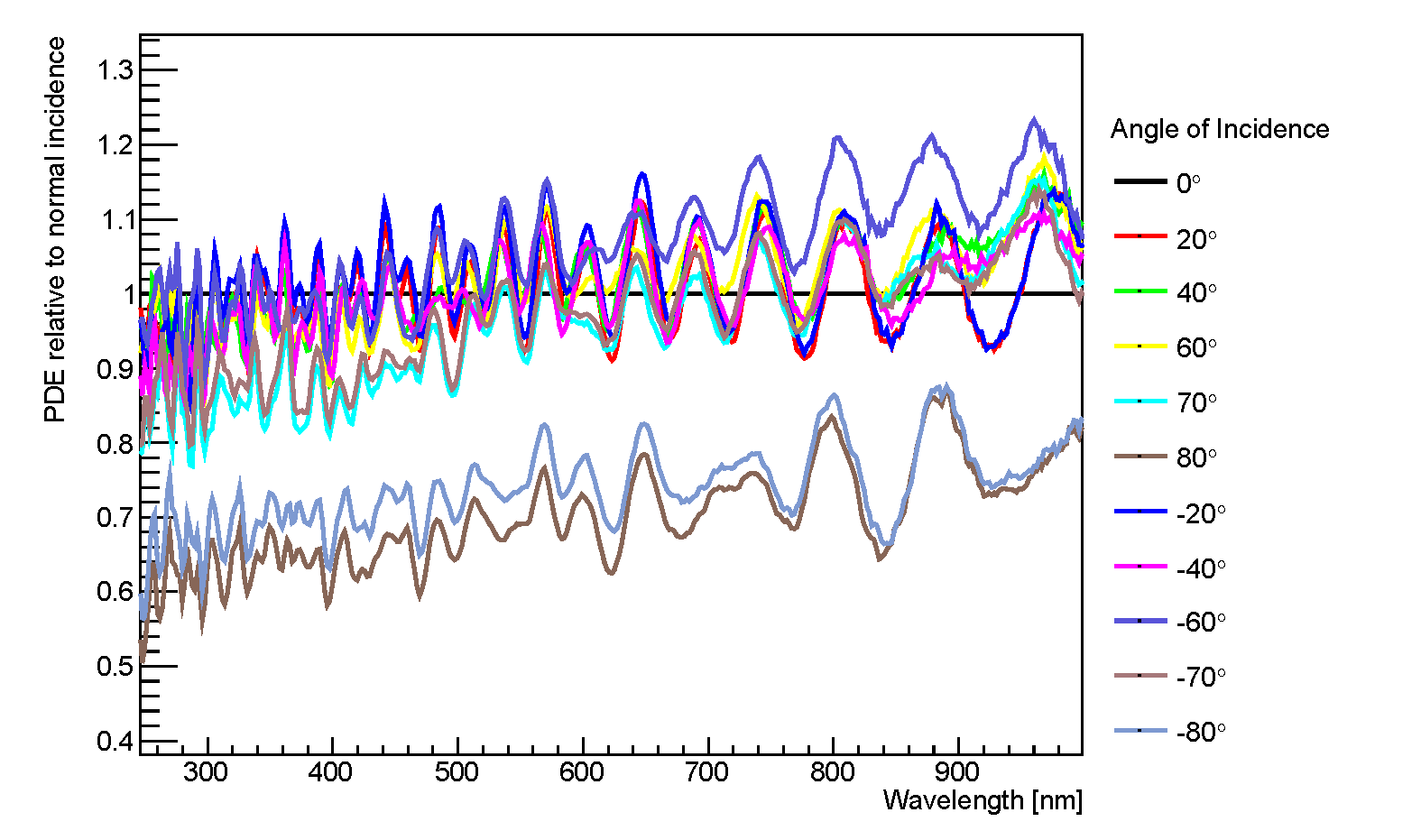} 
    \end{center}
    \caption{\label{fig:relPDEvsAngle} Relative Photon Detection Efficiency of the S14521 vs.\ wavelength for normal up to $80^\circ$ angle of incidence. The difference between positive and negative angles is due to small misalignments of the angle of incidence.}
\end{figure}

Figure \ref{fig:PDEvsAngle} shows the PDE of a S14521 versus wavelength measured at six angles of incidence between $0^\circ$ normal incidence and $80^\circ$. In these measurements, the SiPM was biased 2.25\,V above breakdown, which equates to a 6\% relative overvoltage, which is below our nominal SiPM bias of about 12\% overvoltage for the S14521. However, we find that the normalized spectral response of the S14521 shows little dependence on the bias voltage and thus conclude that the change of the PDE normalized to normal incidence shown in Figure \ref{fig:relPDEvsAngle} is the same for all bias voltages, including the ones we will use in the operation of the \emph{Trinity} and EUSO-SPB2 cameras.

The oscillations in Figure \ref{fig:PDEvsAngle} are due to interference at and below the surface of the SiPM. The oscillations shift to higher and lower wavelengths with changing angle of incidence, resulting in the pattern observed in Figure \ref{fig:relPDEvsAngle}. Within the systematic uncertainties of our measurements, the PDE does not decrease up to angles of $60^\circ$. 

To calibrate the absolute photo response of the camera, we have measured the PDE vs.\ wavelength for one SiPM in each SiPM matrix with light under normal incidence. The remaining SiPMs are calibrated in situ by comparing their response to the calibrated SiPMs when the entire camera is uniformly illuminated with flashes from an LED light pulser. In the in-situ calibration, we make use of the known breakdown voltages and the effective cell capacitances of each SiPM discussed in Section \ref{sec:accepttest} to separate PDE from gain. 

\begin{figure}[!htb]
        \centering
        \includegraphics[width=\columnwidth]{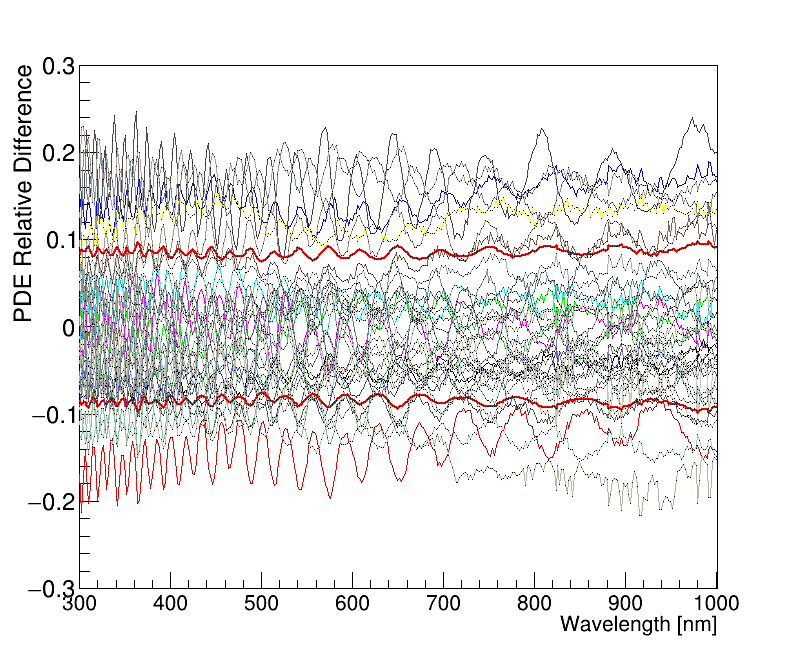}
        \caption{Distribution of the relative photon detection efficiency vs.\ wavelength for 32 of the EUSO-SPB2 camera SiPMs. The standard deviation of the distribution is 9\%, larger than the 2\% statistical uncertainty of one measurement. All SiPMs are biased 4.5\,V above breakdown voltage.}
    \label{fig:PDE_Dist} 
\end{figure}

\begin{figure}[!htb]
    \centering  
     \includegraphics[width=\columnwidth]{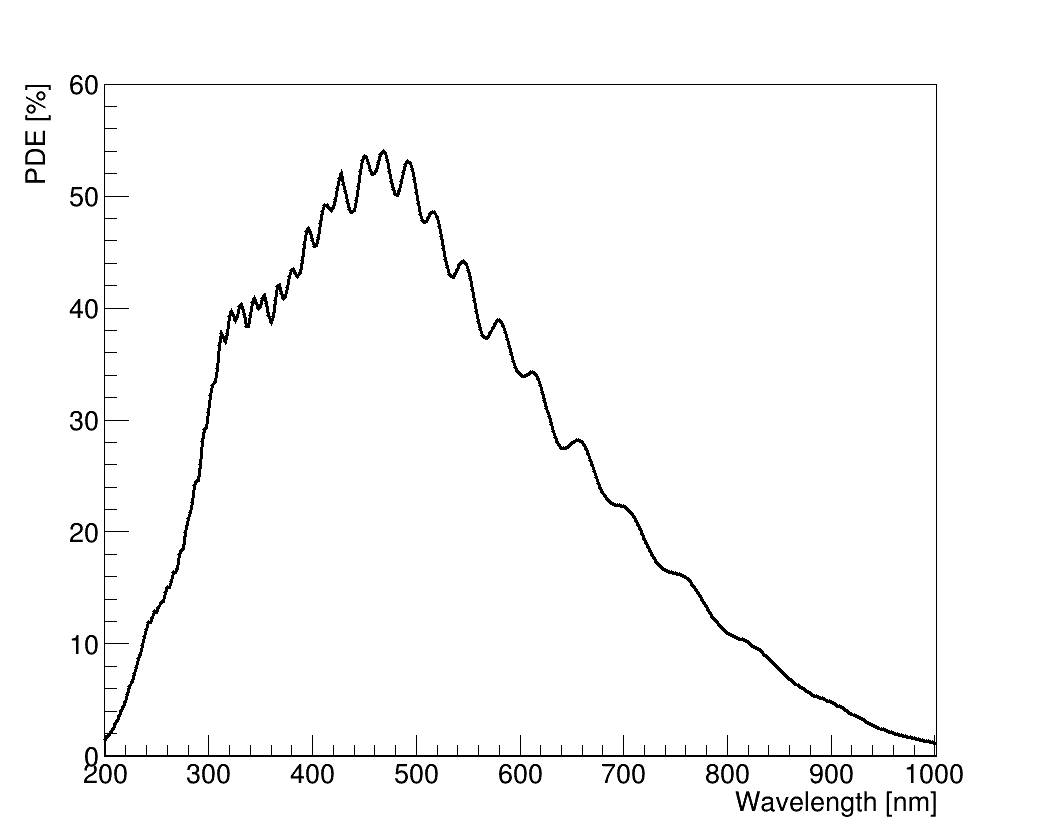}
    \caption{Average PDE calculated from 32 EUSO-SPB2 camera SiPMs. Oscillations in the PDE are from diffraction stemming from the surface layers of the SiPM. The PDE was measured at 4.5\,V overvoltage or at 90\% breakdown probability.}
    \label{fig:Spectral_Response}
\end{figure}

The relative calibration exploits that the spectral response does not change from one SiPM to the next except for a scaling factor, which is illustrated in Figure \ref{fig:PDE_Dist}. The response of all the 32 measured EUSO-SPB2 SiPMs relative to the average of all 32 measurements is flat versus wavelengths. The small oscillations in Figure \ref{fig:PDE_Dist} are due to slight wavelength shifts of the oscillations seen in the spectral response of each SiPM. For comparison, Figure \ref{fig:Spectral_Response} shows the average PDE of all 32 EUSO-SPB2 SiPMs biased at 4.5\,V overvoltage or 12\% above breakdown voltage. The relative response curves of the 32 measured SiPMs vary by 9\% (one standard deviation marked by the red line in Figure \ref{fig:PDE_Dist}) around the average. The statistical uncertainty of the PDE of one SiPM is about 2\%.

\section{Characterization of the Signal Chain and Discriminator\label{sec:signalchain}}

\begin{figure}[!htb]
	\centering      
    \includegraphics[width=\columnwidth]{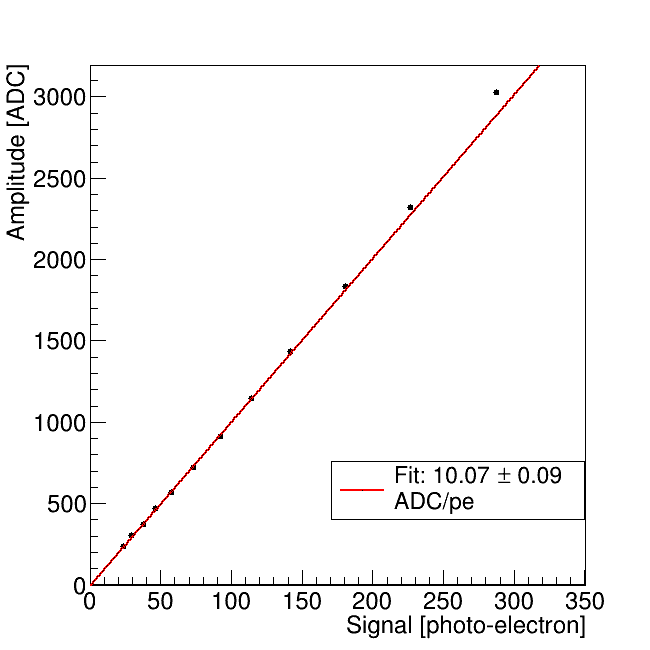}
	\caption{\label{fig:SignalChain}Linearity and dynamic range of the signal chain. The abscissa shows the laser intensity as the average recorded photoelectrons. The red line shows the result of a fit of the data with a linear function.
    }
\end{figure}

\begin{figure}[!htb]
    \centering
    \includegraphics[width=\columnwidth]{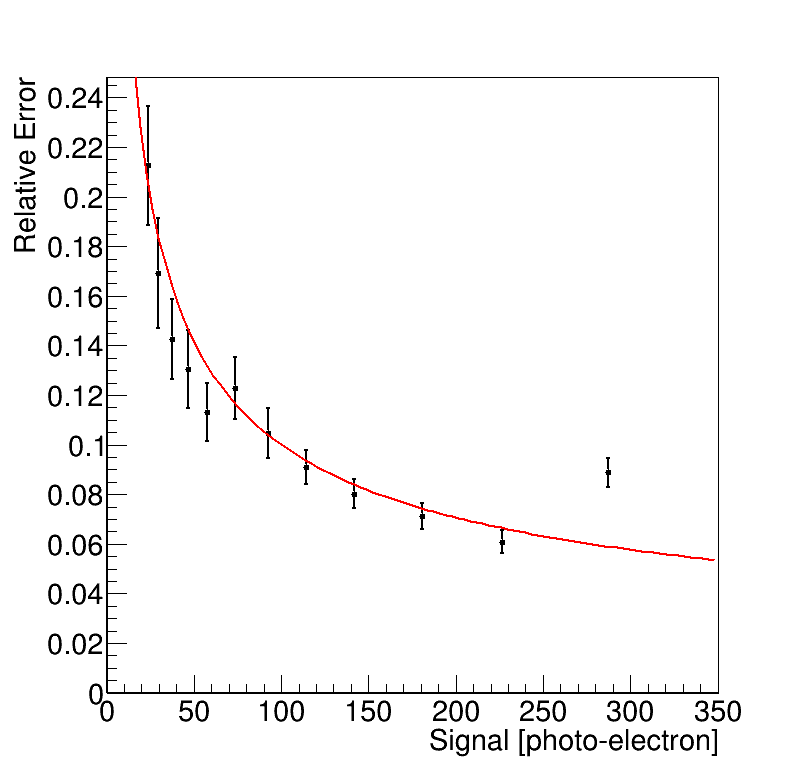}
	\caption{Relative error compared with Poisson limit. \label{fig:RelativeError}}
\end{figure}

\subsection{Linearity of the Signal Chain}

For an absolute calibration of the cameras, we also need to know the conversion factor from the digitized amplitude of the SiPM signals into the number of photoelectrons or photons detected by the SiPM. We obtained that conversion factor and characterized the linearity of the signal chain by flashing the SiPMs with our Picoquant laser and varying the intensity of the light flashes with neutral density filters to scan the entire dynamic range of the digitizer. In these measurements, it is safe to assume that the 120\,ps-wide laser pulse does not add additional width to the 30\,ns Full-Width-at-Half-Maximum (FWHM) pulse shape recorded by the digitizer. The pulse width is mostly determined by the transfer function of the low-pass filter in front of the digitizer. Thus, the normalized pulse shape does not change with the intensity of the light pulse within the linear range of the signal chain and it is also representative of the pulse shape expected for a single photoelectron signal.

A drawback of the 30\,ns wide signals is that they significantly overlap single photoelectron signals from dark counts, thus inhibiting the identification of single photoelectron signals in the digitized traces. In order to obtain an absolute conversion factor based on single-photoelectron signals, we also recorded the signals with a 500\,MHz bandwidth oscilloscope (Tektronix TDS3054C) by tapping into the signal chain before the low-pass filter. From the clearly identifiable single photoelectron signals on the oscilloscope, we thus obtained the conversion factor from signal amplitude into photoelectrons for any signal recorded with the oscilloscope. With that conversion factor, we measured the average number of photoelectrons for each laser intensity. 

Combining the average number of photoelectrons from the oscilloscope measurement with the average digitized amplitude from the AGET system, we then obtain a conversion factor of $10.07\pm0.09$ digital counts per photoelectron for the digitized signal amplitudes in photoelectrons for the EUSO-SPB2 SiPMs. Figure \ref{fig:SignalChain} shows the measurements for different laser intensities. The SiPMs are biased at the nominal 4.5\,V above breakdown in this calibration. We find that the signal chain is linear up to a signal of 300 photoelectrons. For more intense signals, the digitizer frontend saturates, and the response becomes nonlinear.

Figure \ref{fig:RelativeError} shows the relative standard deviation of the recorded amplitudes for each laser intensity between 20 photoelectrons and 300 photoelectrons. In case the pulse-to-pulse fluctuations are only due to Poisson statistics, the expected distribution of the recorded amplitude distribution has a relative standard deviation of $\sqrt{\mu}/\mu$, where $\mu$ is the average number of photoelectrons per pulse and $\sqrt{\mu}$ is the standard deviation of the amplitude distribution in units of photoelectrons. 

The relative width for each intensity agrees well with expectations for Poisson statistics indicated by the red line in the figure. We conclude that noise in the electronics does not significantly contribute to the fluctuations for recorded signals above 20 photoelectrons. For amplitudes below 20 photoelectrons, the fluctuations are increasingly dominated by baseline fluctuations due to dark counts. For signals with 300 or more photoelectrons, we observe that the relative width increases (see the last data point in Figure \ref{fig:RelativeError}) due to the onset of saturation in the signal chain. 

\begin{figure}[!htb]
    \centering
    \includegraphics[width=\columnwidth]{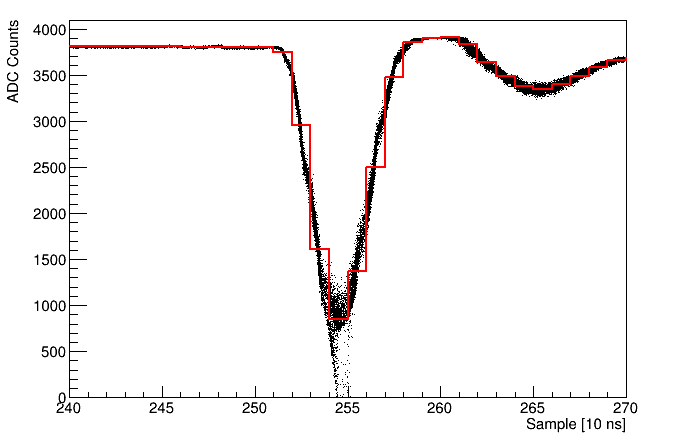}
    \caption{Superposition of the jitter-corrected pulse shapes recorded by flashing a camera pixel with, on average, 288 photoelectrons.}
    \label{fig:SatOnset}
\end{figure}

For each laser intensity, we also measured the average pulse shape, which we are using in our Monte Carlo simulation to model the camera response. When we recorded the laser pulses with the AGET system, we also recorded with the AGET the gate-generator signal that triggered the picosecond laser. In the averaging procedure, we then shifted the recorded trigger pulse relative to the trigger pulse recorded with the first laser pulse until the Chi-square between the two recorded trigger pulses was minimized. We then added the time-shifted samples to the average trace. In this way, we eliminate the phase jitter between the gate-generator signal and the sampling clock of the AGET system and obtain a pulse shape with sub-nanosecond resolution. Figure \ref{fig:SatOnset} shows one example. 

\subsection{Crosstalk between camera pixels}

We characterized the crosstalk between camera pixels by flashing our picosecond laser at a SiPM and recording the response in all 64 channels that connect to the same AGET chip. Unlike in the acceptance testing, where we illuminated the entire SiPM, we placed an adjustable pinhole in front of the SiPM to ensure light does not spill over into neighboring channels. With the pinhole in place and the SiPM biased 4.5\,V above breakdown, we typically detected 150 photoelectrons per laser pulse.

\begin{figure}[!htb]
\centering
        \includegraphics[width=0.9\columnwidth]{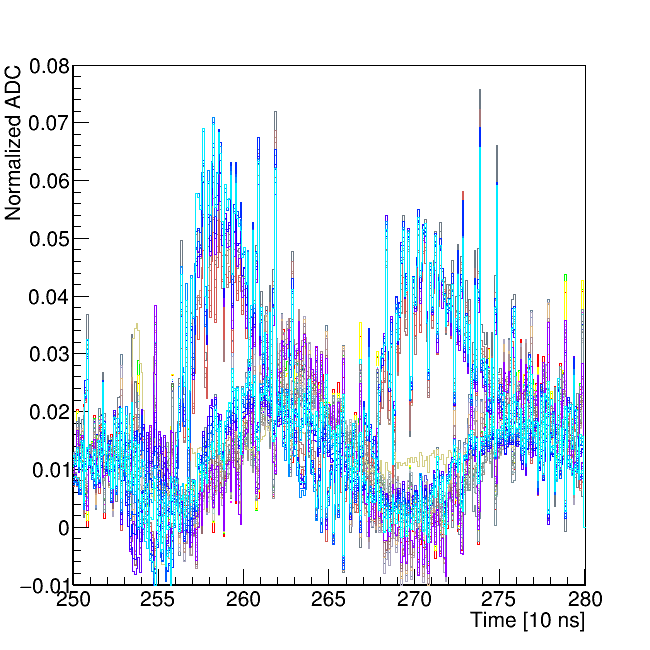}
        \caption{Superposition of average traces of 63 pixels each in a different color. The trace amplitudes are normalized to the amplitude of the signal observed in the pixel flashed with the picosecond laser. The maximum amplitude for each pixel is shown in Figure \ref{fig:CrossTalkCam}.}
        \label{fig:CrossTalkChan}
\end{figure}

\begin{figure}[!htb]
   \centering
    \includegraphics[width=\columnwidth]{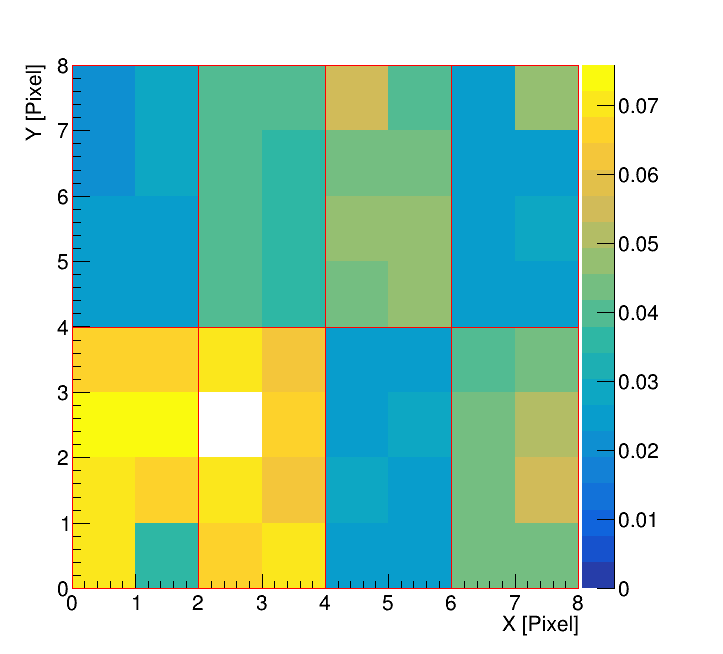}
    \caption{One example crosstalk measurement where the white pixel is illuminated with a picosecond laser. The pixels with the highest crosstalk are connected to the same SIAB as the flashed pixel. All 64 pixels in the figure are connected to the same AGET chip.}
        \label{fig:CrossTalkCam}
\end{figure}

We recorded 1,000 laser pulses for each crosstalk measurement and shifted the recorded traces in time as we did above to obtain one average trace for each pixel. Figure \ref{fig:CrossTalkChan} shows the average traces of all 63 pixels excluding the pixel we flashed with the laser. We observe that the crosstalk signals in all 63 pixels are 30\,ns delayed relative to the laser signal. We quantify the crosstalk in a pixel by recording the maximum amplitude in its average trace and normalizing it to the signal amplitude in the flashed pixel. Figure \ref{fig:CrossTalkCam} gives one example measurement where the white pixel has been flashed. The sixteen pixels in the lower left quadrant belong to the same SIAB and show the highest crosstalk of 7\%. The crosstalk in the remaining channels amounts to 5\% and is due to crosstalk in the low-pass filter boards attached to the digitizer. 

\begin{figure}[!htb]
    \centering
    \includegraphics[width=\columnwidth]{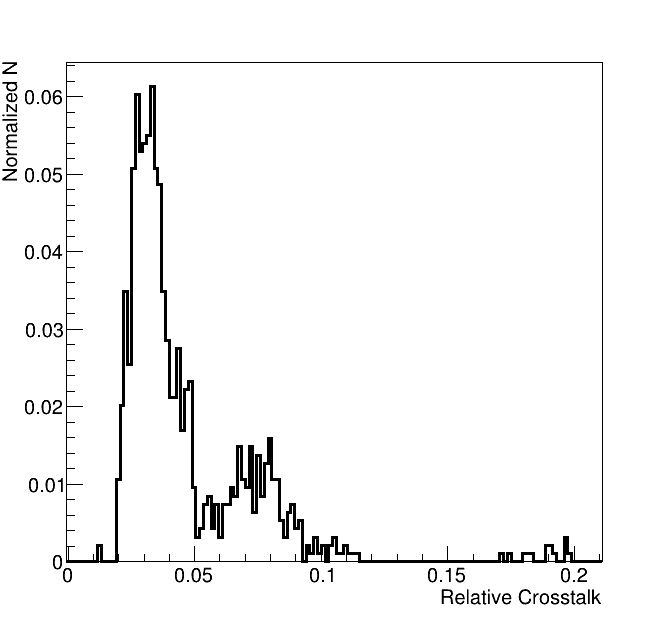}
    \caption{Crosstalk distribution between pixels connected to the same AGET.}
    \label{fig:CrossTalkDist}
\end{figure}

Repeating the measurement by successively flashing all 64 pixels, we arrive at the crosstalk distribution in Figure \ref{fig:CrossTalkDist}. The majority of the crosstalk is below 5\%. The average is 5\%, and in a handful of cases, we observe a crosstalk of almost 20\%.

\subsection{Linearity of the Discriminator}

\begin{figure}[!htb]
    \centering
    \includegraphics[width=\columnwidth]{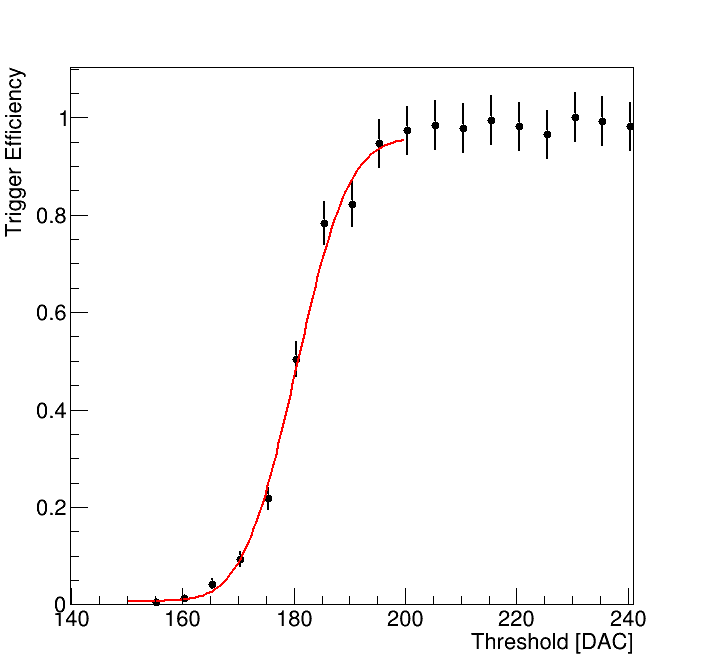}
    \caption{Trigger rate scan for a laser intensity of 88 photoelectrons.}
    \label{fig:TrigEfficiency}
\end{figure}

\begin{figure}[!htb]
    \centering
    \includegraphics[width=\columnwidth]{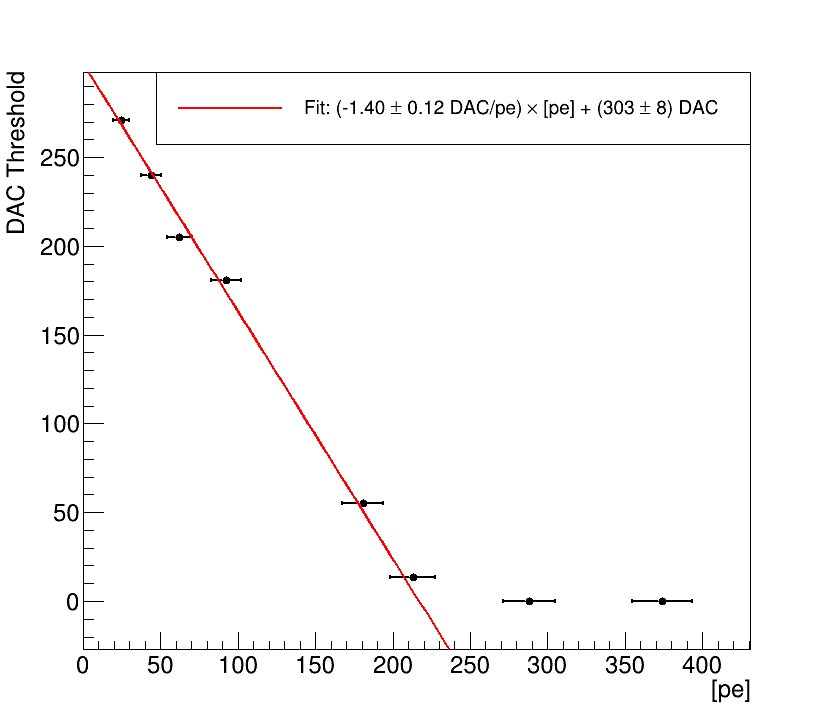}
    \caption{Calibration curve of the eMUSIC discriminator. After about 200 pe the discriminator saturates and the trigger threshold is always exceeded.}
    \label{fig:DiscLin}
\end{figure}

We characterized the linearity of the discriminator and calibrated the discriminator threshold in units of photoelectrons with the same setup we used to measure the linearity of the signal chain. The SiPM bias voltage is set to 4.5\,V above the breakdown voltage. The discriminator threshold is adjustable in 512 steps, where 512 is the lowest and 0 is the highest threshold. We scanned the discriminator threshold for a given laser intensity and recorded the trigger rate for each discriminator setting. Figure \ref{fig:TrigEfficiency} shows the trigger rate versus discriminator setting for one such measurement when the laser intensity was set to 88 photoelectrons. The trigger rate is normalized to the pulse rate of the laser. As the discriminator threshold goes from a small value to a high value (high to low threshold), the trigger rate increases until the discriminator triggers on all laser pulses. We define as threshold the discriminator setting where 50\% of the laser pulses are registered.

Figure \ref{fig:DiscLin} shows the thus-determined discriminator thresholds versus different laser intensities for one camera pixel. From 0 to 200 photoelectrons amplitudes, the discriminator responds linearly before it reaches its highest threshold. We fit the linear range and obtain a calibration factor of $-1.4\pm0.1$ discriminator threshold steps per photoelectron with a positive offset of 303. The discriminators of all camera pixels are calibrated in this way.

\section{Flatfielding\label{sec:flatfielding}}

A uniform camera response simplifies its operation and data analysis. It is achieved by adjusting the SiPM bias such that the product of PDE, SiPM gain, and signal-chain gains and losses is the same for all pixels in the camera. Flat-fielding is the procedure of adjusting the SiPM bias to achieve a uniform camera response.

\begin{figure}[!htb]
    \centering
    \includegraphics[width=\columnwidth]{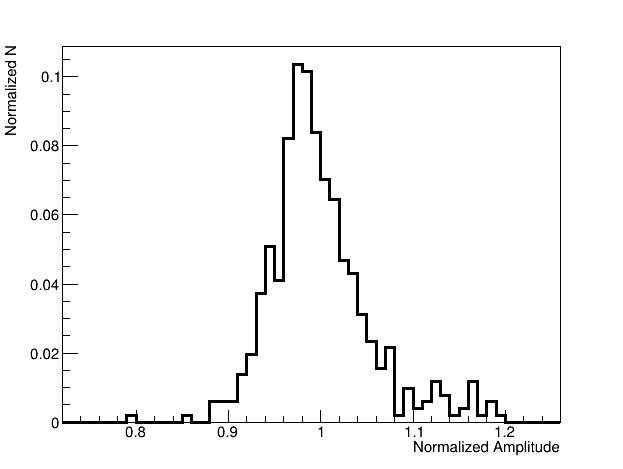}
    \caption{Distribution of the average amplitudes of all camera pixels after flat fielding.}
    \label{fig:FlatFieldDisp}
\end{figure}

In the flat-fielding procedure, the camera is uniformly illuminated with a pulsed light source. For the first round of light flashes, all SiPMs are biased 4.5\,V above their respective breakdown voltage, the bias voltage where the PDE reaches 90\% of its maximum value. From the recorded signals, an average amplitude is calculated per pixel $A_p$, and then the average pixel values are averaged across the entire camera $A_c$. The SiPM bias correction $\Delta U_{\mbox{bias}}$ for a pixel is then obtained by multiplying the nominal 4.5\,V overvoltage with the relative deviation of the pixel average from the camera average. 
\begin{equation}
    \Delta U_{\mbox{bias}} = 4.5\,\mbox{V}\cdot\frac{A_c-A_p}{A_c}
\end{equation}
The SiPM bias correction is then subtracted from the eMUSIC trim voltage for that pixel. After updating the bias voltages of all pixels accordingly, the camera is flashed again with light pulses, and the procedure is repeated until a uniform camera response is achieved. After completion of the flat-fielding, the pixel response is uniform, with a standard deviation of 0.05 (see Figure \ref{fig:FlatFieldDisp}). This is mostly dominated by the finite number of 1,000 flashes, which results in a relative uncertainty of the per-pixel average amplitude of $1/\sqrt{1,000}=0.03$.

\section{Current Monitor\label{sec:currentmonitor}}

We characterized and calibrated the current monitor output of the eMUSICs by illuminating the SiPMs with a steady LED. By varying the current through the LED, we varied its intensity and thus could scan a wide range of SiPM currents from $\sim10\,\mu$A to $450\,\mu$A. For comparison, the typical SiPM current in the \emph{Trinity} Demonstrator is $14\,\mu$A per pixel.

\begin{figure}[!htb]
    \centering
    \includegraphics[width=\columnwidth]{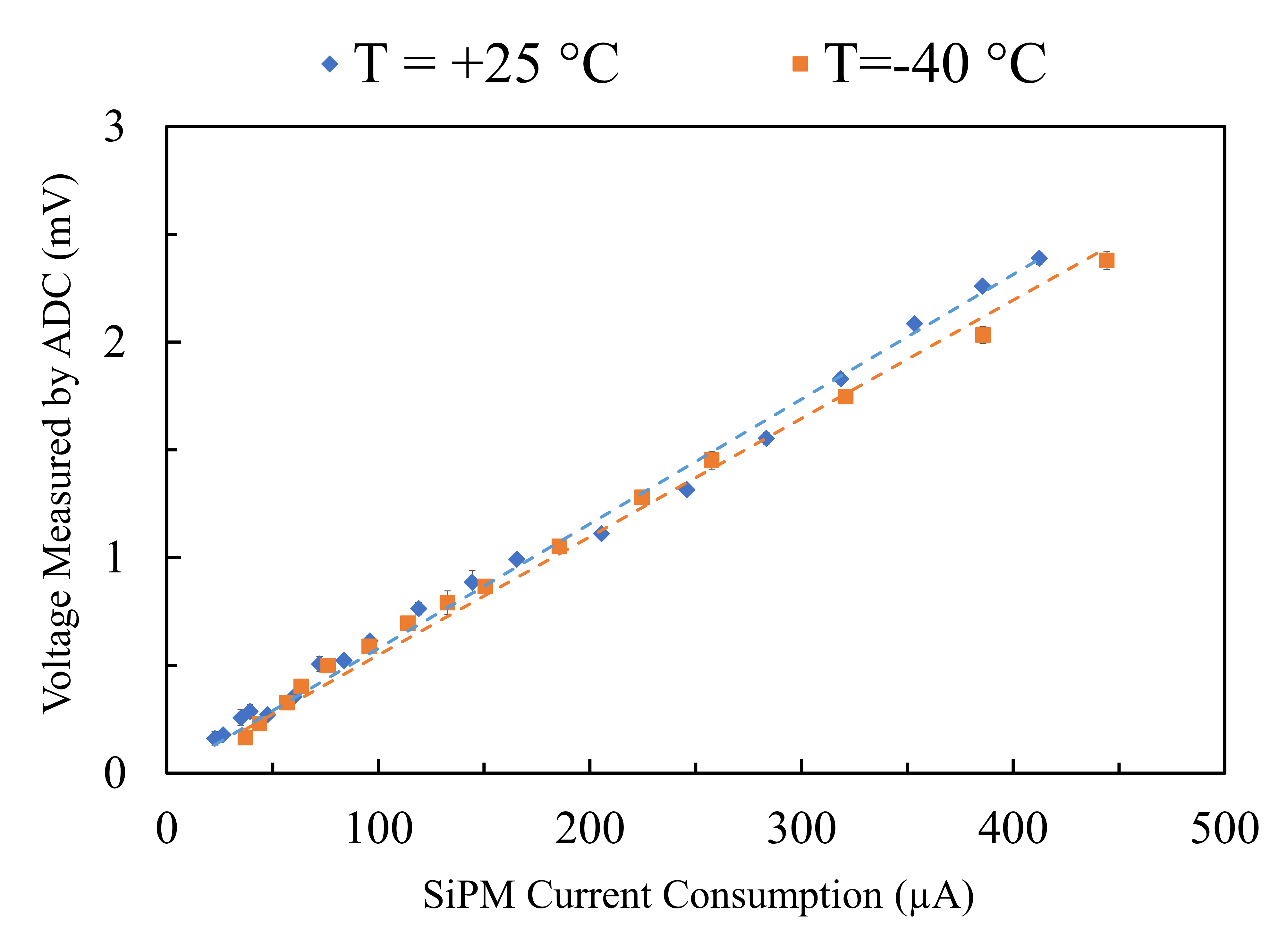}
    \caption{SiPM current monitor measurement. The SiPM is exposed to different amounts of background illumination. }
    \label{fig:CurrentMonitor}
\end{figure}

Figure \ref{fig:CurrentMonitor} shows that the current monitor output is linear over the tested range and shows very little dependence on temperature. We tested the current monitor at $-40^\circ\text{C}$ and $25^\circ\text{C}$. At low SiPM currents, the current monitor is limited by the small $5.3\,\mu\text{V}/\mu\text{A}$ slope such that despite the 16-bit effective resolution of the ADC, we obtain a resolution of $7.6\,\mu\text{A}/\text{ADC count}$.

\section{Summary and Outlook}

We have developed and built the cameras and readout for the \emph{Trinity} Demonstrator and the EUSO-SPB2 Cherenkov telescope. Both are pathfinder and pioneering experiments exploring the detection of Earth-skimming VHE and UHE neutrinos with the imaging atmospheric Cherenkov technique.

Our cameras are the first SiPM cameras aiming at detecting Earth-skimming neutrinos. The spectral response of the SiPMs is a good fit for the Cherenkov spectrum expected to be detected by \emph{Trinity}'s telescopes \cite{Otte2019d}, allowing for compact telescopes with good sensitivity and low energy threshold.

In building and bench testing the cameras, we verified their functionality and characterized and calibrated the photon detection efficiency, the response of the signal chain, the trigger system, and the current monitor. These characterizations provide the necessary parameters to model our cameras in the Monte Carlo simulation of both experiments and evaluate their sensitivity to Earth-skimming neutrinos.

\begin{figure}[!htb]
    \centering
    \includegraphics[trim={1cm 0 2.1cm 0},clip,width=0.8\columnwidth]{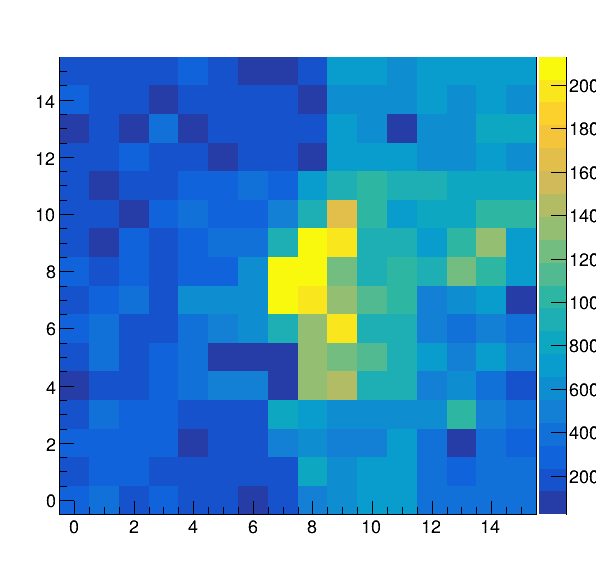}
    \caption{Uncalibrated air-shower image of a cosmic ray recorded with our camera installed in the \emph{Trinity} Demonstrator. The colors indicate the amount of Cherenkov light detected in each pixel.}
    \label{fig:airshowerimage}
\end{figure}

Our instruments meet the requirements of the \emph{Trinity} Demonstrator and EUSO-SPB2 discussed in Section \ref{sec:design} and have been successfully integrated into their telescopes. The EUSO-SPB2 project had its flight in May 2023, and despite an unexpectedly short flight of only 36 hours and 52 min before the balloon plunged into the South Pacific, we could demonstrate that the camera performed as expected \cite{Gazda2023}. The \emph{Trinity} Demonstrator camera had been integrated into its telescope in October 2023, with regular observations starting shortly thereafter. The camera and readout are performing as expected and have been operating without technical issues since. Figure \ref{fig:airshowerimage} shows an air-shower image of a cosmic ray recorded with the \emph{Trinity} Demonstrator during the commissioning phase.

The development of this first-generation SiPM camera for Earth-skimming neutrino detection is only the first step toward a mature design, particularly for \emph{Trinity}. Operating the \emph{Trinity} Demonstrator camera for an extended period will give us the experience of observing Earth-skimming neutrinos from mountaintops and maybe even yield the detection of the first VHE neutrino. With the experience gained by operating the \emph{Trinity} Demonstrator, we will design the next-generation camera, which we plan for the \emph{Trinity} Prototype telescope, which has a $60^\circ$ horizontal field of view and will need to be instrumented with a 3,300-pixel SiPM camera.

\subsection* {Acknowledgments}
We acknowledge the excellent work done by the Georgia Tech Montgomery Machining Mall staff and are grateful for the many collegial and inspiring discussions with our EUSO-SPB2 and \emph{Trinity} collaborators over the past five years. This work was funded with NASA APRA awards 80NSSC19K0627 and 80NSSC22K0426 and funding from the National Science Foundation with award PHY-2112769.


\bibliography{references}   
\bibliographystyle{elsarticle-num-names.bst} 

\end{document}